\documentclass[draftclsnofoot,onecolumn,11pt]{IEEEtran}

\usepackage{amsthm,amssymb,graphicx,graphicx,multirow,amsmath,color,cite}
\newcommand{\ceil}[1]{\lceil #1\rceil}

\def\E{\mathbb{E}}

\def\P{\mathbb{P}}
\def\ie{{\em i.e.}}

\def\R{\mathbb{R}}

\def\Z{\mathbb{Z}}
\def\calN{\mathcal{N}}

\def\BB{\mathbf{V}_\mathtt{a}}
\def\BBB{\mathbf{V}_\mathtt{1}}

\def\i{\mathbf{1}}
\def\pp{\mathsf{\phi}}

\def\d{\mathrm{d}}
\def\k{\mathtt{\sigma}_\eta}
\def\m{\mathtt{\mu_\eta}}

\def\p{\mathtt{P_s}}
\def\pn{\mathtt{P_0}}
\def\x{\mathtt{x}}
\def\y{\mathtt{y}}
\def\z{\mathtt{z}}
\def\I{\mathtt{I}}
\def\pp{\mathtt{P}}
\def\ps{\mathtt{W_{sd}}}

\def\F{\mathtt{F_{sd}}}

\def\a{\mathtt{a}}
\def\sp{\rho^{(2)}_\eta}
\def\som{\mathcal{K}_\eta}

\def\tp{\rho^{(3)}_\eta}
\DeclareMathOperator{\TC}{\mathtt{TC}}
\DeclareMathOperator{\TCL}{\mathtt{TC_l}}
\DeclareMathOperator{\TCU}{\mathtt{TC_u}}

\def\l{\ell}
  
  \newcommand{\h}[1]{\ensuremath{\mathtt{h}_{#1}}}
  
\def\PP{\mathbb{P}^{!o}}

\def\EP{\mathbb{E}^{!o}}
\def\T{\theta}
\def\sinr{\mathtt{SINR}}

\def\sir{\mathtt{SIR}}
\def\W{\mathtt{N}}

\newtheorem{lemma}{Lemma}{}
  \newtheorem{thm}{Theorem}
  \newtheorem{theorem}{Theorem}
  
  \newtheorem{cor}[thm]{Corollary}

  \newtheorem{definition}[thm]{Definition}

\def\ch#1{#1}
\def\chr#1{#1}

\begin{document}
\title{High-SIR Transmission Capacity of Wireless Networks with General Fading
and Node Distribution}
\author{Radha Krishna Ganti, Jeffrey G. Andrews, and Martin Haenggi
\thanks{R. K. Ganti and J. G. Andrews are with the University of Texas at
Austin, TX, USA,
and M. Haenggi is with the University of Notre Dame, IN, USA. The contact author
is R. K. Ganti, {\tt rganti@austin.utexas.edu}. Date revised: \today}}

%
\maketitle
\begin{abstract}
\chr{
In many wireless systems, interference is the main performance-limiting factor, and is primarily dictated by the locations of concurrent transmitters.  In many earlier works, the locations of the transmitters is often modeled as a Poisson point process for analytical tractability.
 While analytically convenient, the PPP only accurately models networks whose nodes are placed independently {\em and} use ALOHA as the channel access protocol, which preserves
the independence. 
Correlations between transmitter locations  in non-Poisson networks,  which model intelligent  access protocols, makes the outage analysis extremely difficult.}
In this paper, we take an alternative approach and focus on an asymptotic regime where the density of interferers $\eta$ goes to $0$. We prove for general node distributions and
fading statistics that the success probability $\p \sim 1-\gamma \eta^{\kappa}$
for $\eta \rightarrow 0$, and provide values of $\gamma$ and $\kappa$ for a
number of important special cases. We show that $\kappa$ is lower bounded by $1$
and upper bounded by a value that depends on the path loss exponent and the
fading. This new analytical framework is then used to characterize the
transmission capacity of a very general class of networks, defined as the
maximum spatial density of active links given an outage constraint.
\end{abstract}

\section{Introduction}
Network information theory attempts to characterize fundamental limits on
networked systems of nodes, in particular the rate at which information can
propagate through the network between an arbitrary set of nodes.  In a wireless
network, the first order effect governing the flow of information is the spatial
separation between the nodes, which determines the desired signal and
interference power at each receiver.  Since \chr{for large networks,} most known information-theoretic approaches lead either to  loose upper bounds or
are typically lacking in tractability, a considerable number of alternative
approaches that lead to characterizations of basic properties of wireless
networks are under active investigation.

Considerable recent interest has been paid to the impact that the user geometry
-- i.e., the locations of the transmitters and receivers -- has on fundamental
metrics of wireless network performance \cite{net:Haenggi09jsac, BaccelliNOW}.
Since it is impractical (and in most cases, impossible) to enumerate all such
locations, it is typical to model their coordinates as drawn from a random
two-dimensional point process. The most commonly used spatial random process is
the homogeneous Poisson point process (PPP), which is by far the most tractable
and well understood, as it presumes independence between each and every node
location, as well as a homogeneous distribution over space.   Even if the node
locations are in fact independent and uniformly scattered in space, most
``good'' scheduling algorithms  \chr{\ch{induce} correlations in the  transmitter  locations} to avoid collisions.  This has led to a sharp
tradeoff between tractability -- requiring essentially an interference-agnostic
ALOHA multiple access control (MAC) protocol that preserves the independence --
and optimality/generality, whereby more intelligent MAC protocols can be
characterized despite introducing dependence and therefore requiring a more
general point process.

The objective of this paper is \chr{to} provide a new mathematical framework for
\chr{analyzing}  a very general class of wireless networks.  In particular, this
paper provides novel tools for modeling interference, computing outage and
transmission capacity for a general spatial distribution of transmitters, and under
a general fading model.

\subsection{ Background and Related Work}

Among the first papers to use a PPP to model the node locations in a  wireless
network is \cite{Sousa90} where the outage probability of $n$ users uniformly
distributed in a bounded area $A\subset \R^2$ was found with no channel fading,
by scaling $A$ and $n$ appropriately. The interference distribution in a Poisson
network was shown \cite{ilow1998aas} to be a stable distribution with parameter
$2/\alpha$, where $\alpha$ is the path loss exponent. Both these results as well
as subsequent analysis of Poisson networks hinge mainly on two mathematical
tools: the probability generating functional (PGFL) and the Palm
characterization. Both of these are fairly simple for a PPP because of the
independence between the node locations. The conditional PGFL is known only for
a PPP (and a few variants), and this has made it extremely difficult to extend
the rich set of results on PPPs to other more general spatial node
distributions.

In the domain of PPP-modeled networks, significant recent work has developed
sophisticated mathematical models on the interference experienced in such
networks
\cite{BacZuy97,BacKle97,zorzi1995optimum,hellebrandt1997cip,HaenggiNOW,
jagadish06allerton,chan2001calculating,WinPin09,jsac:baccelli}.  In such cases
it is then possible to compute the probability that a typical receiver has a
signal-to-noise-interference ratio below some threshold $\theta$, which is
termed the \emph{outage probability} \cite{bacelli-aloha,WebAndJin07}.  A final
related metric is the \emph{transmission capacity} (TC), which is the maximum
number of successful transmissions per unit area at a specified target outage
probability \cite{weber:2005,WebAndJinTut}. \chr{ In many of these early works,}   interference is typically
treated as noise, \chr{but this assumption can be relaxed}. 
Transmit/receive approaches that are known to be optimal in some multiuser
settings, such as dirty-paper coding \cite{KouAnd09a} or successive interference
cancellation \cite{WebAnd07,huang-spatial,vaze2009transmission}, can be
considered.  In this paper, however,
interference is treated as noise.

There has been significant interest in moving beyond the Poisson process, but
progress has been slow, and is limited to either a deterministic arrangement of
nodes (degenerate point processes) or variants of a Poisson point process.  In
\cite{ganti-2007}, the outage and TC are obtained when the nodes are
distributed as a Poisson cluster process, which results from independent
clustering applied to a Poisson point process.  Carrier sense multiple access
(CSMA) MAC protocol results in networks, where the concurrent transmitters have
a minimum distance, and this precludes PPP as a model for transmitter locations.
Lattice networks are commonly used to model CSMA networks \cite{durvy2009self},
and the outage probability was analyzed in
\cite{silvester, ferrari,Haenggi09twc,mathar1995dci}. Analyzing CSMA protocol
with a
random node distribution is extremely difficult and usually the interfering
transmitters are approximated by a Poisson point process.  In
\cite{BaccelliNOW}, the outage probability was obtained by
approximating
interferer locations in a CSMA protocol, with a non-homogeneous Poisson point
process and  in \cite{HasAnd07,kaynia2008performance,kaynia-joint} by
excluding interferers in guard zone around the typical transmitter in a Poisson
model.

\subsection{Overview of Contributions and Organization}

Since obtaining  outage probability (or TC) for a general class of node
distributions is extremely difficult, we concentrate on the low-interference
regime, which  \chr{corresponds to a low-density of concurrent transmitters}.  For any spatial distribution of the transmitters, we show how to obtain
two constants $\gamma$ and $\kappa$, so that we can characterize the success
probability as
\begin{equation}\p \sim 1- \gamma \eta^{\kappa}, \quad \eta \rightarrow 0,
  \label{eq:top}
\end{equation}
where $1-\p$ is the outage probability of a typical link with the $\sir$ model and $\eta$ is the
density of concurrent transmitters\footnote{\chr{By $f(x)\sim g(x)$, $x\to 0$ we mean that $\lim_{x\to 0} \frac{f(x)}{g(x)} =1$.}}.  
If the two constants $\gamma$ and $\kappa$
can be calculated, then the TC defined in \eqref{eq:TC} is asymptotically
\begin{equation}
  \TC(\epsilon) \sim
\left(\frac{\epsilon}{\gamma}\right)^{1/\kappa}(1-\epsilon),
  \quad \epsilon \rightarrow 0,
  \label{eq:tc_main}
\end{equation}
where $\epsilon$ represents the constraint on the outage probability.  By the
continuity of $\p$, the above approximation will be accurate for $\eta$ close to
zero.

Unlike the exact characterization of the outage probability, obtaining $\gamma$
and $\kappa$ turns out to be possible for a wide range of MAC protocols and node
distributions.   In \cite{ricardo-10,giacomelli-asymptotic,ganti:isit2010}, the outage probability was also approximated in a similar spirit but only
for Rayleigh fading, where it was shown that $\kappa=1$ for any ALOHA network and
$\kappa=\alpha/2$ for CSMA, where $\alpha$ is the path loss exponent.  This
paper however proves such a result under a general fading model, and provides
much more general tools for future non-Poisson network analysis. The
transmission capacity is also derived, and compared to prior results, which are
nearly exclusively for a PPP. More precisely, this paper makes the following
four primary contributions.
\begin{enumerate}
  \item For any plausible MAC protocol, the interference scaling exponent $\kappa$
is shown to be between 1 and $\alpha\nu/2$, where $\alpha$ is the path loss
exponent, and  $\nu \geq 1$ depends only on the fading distribution of a typical link.

  \item For any general fading and underlying node distribution with ALOHA, it is shown that $\kappa=1$, and $\gamma$ is
shown to depend only on the second-order
	product density of the spatial distribution of nodes.  The TC is then
calculated, \chr{and is shown to linearly scale with the outage constraint $\epsilon$.}

  \item The asymptotic outage and the TC for a CSMA network modeled by a Matern
	hard-core point process with a general fading distribution is obtained,
in which $\kappa= \alpha\nu/2$.  From \eqref{eq:tc_main}, the TC of the  CSMA
network is observed to be $\Theta(\epsilon^{\frac{2}{\alpha\nu}})$ which is a
significant improvement compared over  ALOHA networks whose TC is $\Theta(\epsilon)$.

\item The asymptotic TC of any spatial distribution of transmitting nodes
  with Rayleigh fading is obtained and  shown to depend only on the second-order product density of the transmit point
  process.
\end{enumerate}

Finally, we note that although the focus here is on outage and transmission
capacity, the techniques are general and can be extended to other SINR-related
metrics like expected forward progress \cite{bacelli-aloha}.

The paper is organized as follows. In Section \ref{sec:sys_model},
we introduce the system model, the assumptions on the source-destination
distribution and the transmitter point process. In Section \ref{sec:suc_prob},
we
obtain the bounds on $\kappa$.  ALOHA networks are analyzed in Section
\ref{sec:ALOHA} where we show that $\kappa=1$ and provide the exact value of
$\gamma$;  CSMA networks are studied in Section \ref{sec:CSMA}. In  Section \ref{sec:noise}, we consider the asymptotic outage  results of ALOHA and CSMA networks  with thermal noise.
In Section \ref{sec:TC}, we derive the transmission capacity for ALOHA, CSMA
networks, and a
general class of networks with Rayleigh fading. The notation and symbols used
are summarized in Table \ref{tab:one}.

\begin{table}
  \centering
  \begin{tabular}{|l|l|}
	\hline
	Symbol&Description\\ \hline
	$\Phi$& point process representing potential transmitters\\
	$\Phi_t\subset \Phi$& transmitters selected by the MAC protocol \\
	$\eta$& density of transmitters $\Phi_t$\\
	$\rho^{(2)}(x)$& second-order product density of $\Phi$\\
	$\rho^{(k)}_\eta(x_1,\dots,x_{k-1})$&  $k$-th order product density of
	$\Phi_t$\\
	$\PP(A)$ & reduced Palm measure of an event $A$\\
	$\EP$& expectation w.r.t. reduced Palm measure\\
	$\T$ & $\sir$  threshold\\
	$\p$ & success probability, \ie, $\PP(\sir(o,r(o))>\T)$\\
	$\l(x)$& large-scale path loss function\\
	$\alpha$& path loss exponent\\
    $R$& source-destination (link) distance\\
	$\ps$& (power) fading between  source and destination\\
	$\F(x)$& CCDF of $\ps$, \ie, $\P(\ps>x)$\\
	\multirow{2}{*}	{$\nu$}& Exponent of  $x$ of the first term \\
	&of the Taylor series of
	$1-\F(x)$\\
	$\h{}$& (power) fading between an interferer and the receiver\\
	$\I$& interference\\
    $\|x\|$& norm of $x$\\
	$\mathcal{L}_X(s)$& Laplace transform of the random variable $X$\\
	$o$ & origin\\
	$B(x,y)$ & ball of radius $y$ centered around $x$\\
	\hline
  \end{tabular}
  \caption{Notation and symbols used in this paper}
  \label{tab:one}
\end{table}

\section{System Model and Metrics}
\label{sec:sys_model}
We model the location of the potential transmitters by a stationary and
isotropic \cite{stoyan,verejones} point process
$\Phi \subset \R^2$ of unit density on the plane.
 We assume that the MAC protocol
schedules a transmitter set $\Phi_t \subset \Phi$ with the following properties:
\begin{enumerate}
 \item The transmitter set $\Phi_t$ is stationary and isotropic.
 \item The density of the transmitter set is $\eta \leq 1$, and the MAC protocol
can reduce the density $\eta$ to zero.
\end{enumerate}

The large scale path loss model is
denoted by $\l(x): \R^2 \rightarrow [0,\infty]$ and is assumed to have the
following properties:
\begin{enumerate}
 \item $\l(x)$ is a  non-increasing function of $\|x\|$.
 \item $\int_{B(o,\epsilon)^c} \l(x) \d x < \infty,$ for all $\epsilon >0$,
\end{enumerate}
where $B(o,\epsilon)$ denotes the ball of radius $\epsilon$ around the origin.
In this paper we concentrate on the following two models, but the analysis can
be easily extended to other path loss models:
\begin{enumerate}
  \item Bounded path loss model: $\l(x) =(1+\|x\|^{\alpha})^{-1}$, $\alpha>2$.
  \item Non-bounded path loss model: $\l(x)=\|x\|^{-\alpha}$, $\alpha>2$.
\end{enumerate}
For many proofs, the exact form of the path loss model does not matter as long as
$\l(x)=\Theta(\|x\|^{-\alpha})$, $\|x\|\rightarrow \infty$.
\subsection{Success Probability}
We pick a typical transmitter from $\Phi_t$ and add a new receiver (a test
probe to measure interference and  outage) in a random direction at a distance $R$. Since the
process $\Phi_t$ is stationary, we can move the typical transmitter to the
origin, and we denote its receiver by $r(o)$.
The success probability of a typical link is given by
\begin{equation}
 \p= \PP(\sir(o,r(o)) >\T),
\end{equation}
where
\[\sir(o,r(o))= \frac{\ps\l(R)}{\sum_{\z \in
\Phi_t}\h{\z r(o)}\l(\z -r(o))}.\]
$\PP$ represents the reduced Palm probability \cite{stoyan} of $\Phi_t$, and is
equivalent to conditional probability of real-valued random variables.  
\ch{Essentially, $\PP$ is the conditional probability measure of the point process given that there is a point at the origin, but disregarding that
point in all calculations.}
 \chr{For the stationary process $\Phi_t$, the Palm probability of an event $A$ can be interpreted as
\[\PP(A) = \frac{1}{\eta \pi R^2}\E\sum_{x\in \Phi_t\cap B(o,R)} \i( (\Phi_{t}+x)\setminus \{x\} \in A),\]
where $\Phi_t+x$ denotes the translation of $\Phi_t$ by $x$.}
We require the reduced Palm measure 
since we are considering the point process distribution from the point of view of the typical point at the origin. The received power (after appropriate signal processing) normalized by the
path loss $\l(R)$ is denoted by $\ps$, and  $\h{\z r(o)}$ represents the
interference power (normalized by the path loss $\l(\z-r(o))$) of a transmitter
$\z$ at the receiver $r(o)$  after signal processing. We assume that the random
variables $\h{\z r(o)}$ are i.i.d., and
independent of $\ps$.
In this paper, the major emphasis is on $\sir$ rather than $\sinr$, since interference is a major debilitating factor in wireless networks, while noise can be combated by increasing the transmit power to the maximum permissible value. As we shall see in Section \ref{sec:noise}, it turns out that the results for outage considering noise can   easily be obtained using the $\sir$ based outage results.

The success probability can be rewritten as
\begin{eqnarray}
 \p &=& \PP\left(\ps > \frac{\T}{\l(R)} \sum_{\z \in
 \Phi_t}\h{\z r(o)}\l(\z-r(o))\right)\nonumber\\
 &=&\EP \F\left(\frac{\T\I}{\l(R)}\right) ,
 \label{eq:error}
\end{eqnarray}
where
\[\I =\sum_{\z \in \Phi_t}\h{\z r(o)}\l(\z-r(o)),\]
denotes the interference and $\F(x)$ the CCDF of $\ps$.
Evaluating $\EP \F\left(\frac{\T\I}{\l(R)}\right)$ is extremely hard and may
not be possible in most cases. For a general $\F$,  evaluating $\EP \F\left(\frac{\T\I}{\l(R)}\right)$ is not
easy even when $\Phi_t$ is a Poisson point process. To gain insight about the system performance for general MAC protocols and physical layer
technologies, we consider the asymptotic regime $\eta\rightarrow 0$. More
precisely, we try to find two constants, the {\em interference scaling exponent}
$\kappa$  and  the {\em spatial contention parameter} $\gamma$, such that
\begin{equation}
  \p \sim 1-\gamma \eta^{\kappa}, \quad \eta \rightarrow 0.
  \label{eq:asymp}
 \end{equation}
If the two constants $\gamma$ and $\kappa$  are determined exactly, then
\eqref{eq:asymp} is
a good approximation of the success probability near $\eta=0$. In a
wireless network, $\eta\rightarrow0$ corresponds to a high-$\sir$ regime.
 \subsection{Transmission Capacity}
 The transmission capacity (TC) metric was introduced in \cite{weber:2005} to
 quantify the maximum spatial density of simultaneous transmissions
 possible for a given outage constraint, and  it is given by
 \begin{equation}
   \TC(\epsilon)  = (1-\epsilon)\sup\{\eta: \p > 1-\epsilon \}.
   \label{eq:TC}
 \end{equation}
 In this paper we  analyze the transmission capacity in the high-$\sir$ regime when
 \begin{enumerate}
   \item $\F(x)=\exp(- x)$, corresponding to Rayleigh fading,  for any MAC protocol or any point process
 	$\Phi_t$.
   \item  $\F(x)$ is not exponential. Here we restrict the MAC protocol to
ALOHA  	and CSMA.
 \end{enumerate}

\subsection{Received Power Distribution}
The received power $\ps$ depends on the small-scale fading, the diversity order
of the channel, the transmit power and the signal processing at the receiver. Its complimentary cumulative distribution function (CCDF) is denoted by $\F(x)$. A few illustrative examples of $\ps$ are:
\begin{itemize}
 \item If suppose all the nodes transmit at a constant power, then
   $\ps =   \h{o r(o)}$ and corresponds to the small-scale fading.
   If the fading is Rayleigh, $\ps$ is
   exponentially distributed and when the fading is Nakagami-m, $\ps$ is
   gamma distributed.
 \item The nodes in the system may use multiple antennas for communication,
   in which case the distribution of $\ps$ depends on the specific space-time
   technique used.
   If suppose the typical receiver $r(o)$ has $M_r$ receive antennas and
   uses  receive beamforming, then $\ps$ is  $\chi^2$
distributed with $2M_r$ degrees of freedom and mean $M_r$.
 \end{itemize}
To facilitate the analysis, we assume that the CCDF $\F(x)$ satisfies the following
conditions.
\begin{description}
  \item[A.1] The CCDF is analytic over the entire real line, and has a Taylor series
	expansion given by
	\begin{equation}
	  \F(x)= 1-c_o x^{\nu}+\sum_{k=1}^{\infty}\frac{c_k}{k!}x^{k+\nu}, \quad
  c_o\neq 0,
	  \label{eq:ccdf}
	\end{equation}
	where $\nu \in \mathbb{N}$. The coefficients $c_k$ should  satisfy 
	\[\sum_k |c_k| b^{-k} <\infty,\]
	for some $b>1$.
  \item [A.2] The CCDF is Lipschitz, \ie, there exists a $C>0$ such that for
	$x,y>0$
	\[|\F(x)-\F(y)| < C|x-y|.\]
  \item [A.3] $\E[\ps] <\infty$, \ie, $\int_0^\infty \F(x)\d x<\infty$.
  \end{description}
The above conditions are satisfied by many distributions of interest. For
example, if the source-destination fading is Nakagami-m distributed, then
\[\F(x) = \frac{\Gamma(m,mx)}{\Gamma(m)}= 1-
\frac{1}{\Gamma(m)}\sum_{k=0}^{\infty} \frac{(-1)^{k}}{k!}\frac{x^{m+k}}{m+k}.\]
In this case $\nu=m$ and $c_k=(-1)^{k+1}/(m+k)$, which satisfies Condition A.1,
for any $b\geq 1$. For Rayleigh
fading, $\F(x)= \exp(-x)$ and $c_k=(-1)^{k+1}$.
In all the above examples, the derivative of $\F(x)$ is
bounded and continuous, \ie,  $\sup \frac{\d \F(x)}{\d x}$ is bounded and hence $\F(x)$ is
Lipschitz with a constant $\|\frac{\d \F(x)}{\d x}\|_\infty$.
\subsection{Power Distribution of an Interferer}
The power (after normalizing by the path loss) from the interferer $\z \in
\Phi_t$ to the
receiver $r(o)$ is denoted by $\h{\z r(o)}$.   We assume that the random
variables $\h{\z r(o)}$ are i.i.d. and independent of $\ps$, and  we use $\h{}$ to
denote a generic random variable with the distribution of any $\h{\z r(o)}$. We also assume
\begin{description}
  \item[A.4] The moments of $\h{}$  satisfy $\E[\h{}^n] =O( b
	^{n}n!)$, $ n\rightarrow \infty$,
	for some $b>0$.
\end{description}
For example, in the case of a single-antenna system without power control,
$\h{}$ is distributed as the fading (power) coefficient.  In the case of Nakagami-m fading we have
$\E[\h{}^n]= m^{-n}\Gamma(n+m)$,
which satisfies Condition A.4.

\subsection{Point Processes}
In this subsection, we briefly introduce the notation and definitions related
to point processes that we will be using throughout the paper. A detailed
treatment can be found in \cite{stoyan, verejones, cox,snyder}.

\begin{definition}[Second-order product density]
For a stationary and isotropic point process $\Phi_t$ of density $\eta$, the
second-order product density $\sp (x)$ is defined by
the following relation:
\[\EP\sum_{\x\in \Phi_t}\i(\x\in B) = \eta^{-1}\int_B \sp(x)\d x ,\]
where $B$ is any Borel subset of $\R^2$.
\end{definition}
\noindent Intuitively the second-order product density $\sp(x)$  denotes the
probability of finding a pair of points of $\Phi_t$ separated exactly by a distance
$\|x\|$. For a homogeneous PPP of density $\eta$, the second order product density is $\sp(x)=\eta^2$.
\begin{definition}[Second-order reduced moment measure]
  \label{def:secorder}
For a stationary and isotropic point process $\Phi_t$ of density $\eta$, the
second-order reduced moment measure,  denoted by $\som(B), B\subset \R^2$, is
defined as
\[ \som(B)= \eta^{-1}\EP\sum_{\x\in \Phi_t}\i(\x\in B). \]
\end{definition}
\noindent From the definitions of $\som(B)$ and $\sp(x)$, it follows that
\begin{equation}
 \som(B) =\eta^{-2}\int_B \sp(x)\d x.
\end{equation}
The second-order reduced moment measure is a positive and positive-definite
measure \cite{verejones}. Hence for any compact set $B \subset \R^2$, there
exists a constant $C_B$ such that
\begin{equation}
\som(B+x)< C_B,\quad \forall x\in \R^2.
 \label{eq:ppd}
\end{equation}
The Rippley's $K$-function \cite{stoyan} is a special case of the second-order
reduced moment measure,
and is defined as $K_\eta(r)=\mathcal{K}_\eta(B(o,R))$.
\noindent We now begin with the analysis of the exponent $\kappa$.
\section{Bounds on the Interference Scaling Exponent}
\label{sec:suc_prob}
\begin{figure}[ht]
  \begin{center}
	 \includegraphics[width=0.5\columnwidth]{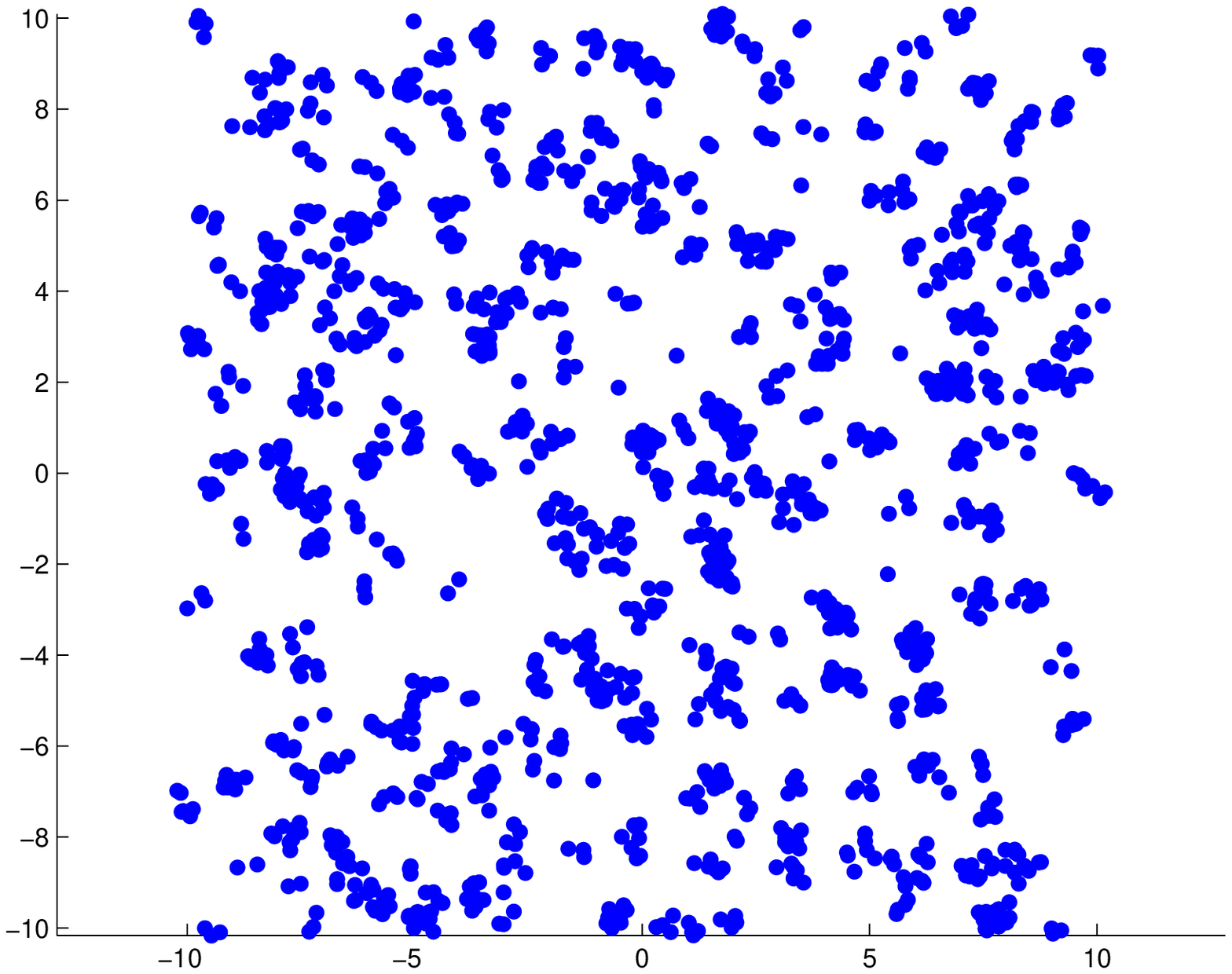}\includegraphics[width=0.5\columnwidth]{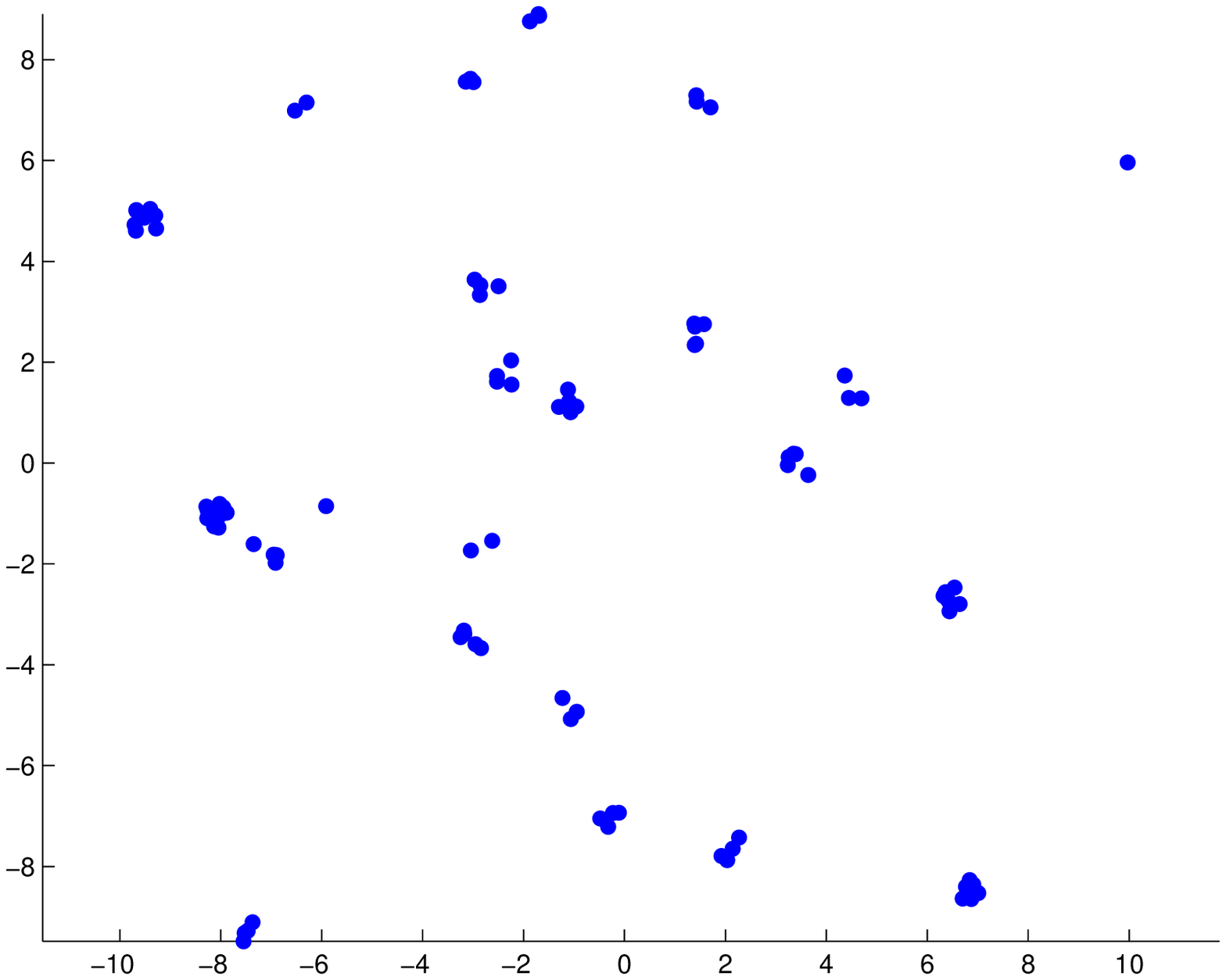}
  \end{center}
  \caption{Illustration of MAC scheme which violates Condition $B.1$. Left: A
  network $\Phi$ modeled by a cluster process with a unit density parent process
  and cluster density $15$.  Right: The MAC protocol selects a cluster (not
  individual nodes) to
  transmit with probability $1/n$, \ie, $n=15$ resulting in
  a cluster process $\Phi_t$ of density $1/15$. We observe
  that even for $n\rightarrow \infty$, the success probability never
  approaches one because of intra-cluster interference.}
  \label{fig:unreasonable}
\end{figure}

At low transmitter density, it is mainly the interference scaling exponent that determines 
the outage, and a higher exponent $\kappa$ implies better outage performance.
From \eqref{eq:error} and \eqref{eq:asymp} it follows
that
\[ \frac{\EP[\F(0)-\F(\I)]}{\eta} = \eta^{\kappa-1} \gamma  + o(\eta^{\kappa-1}).\]
As $\eta \rightarrow 0$, it is intuitive that the interference $\I$ {\em becomes small}, and hence
we can observe that $\kappa$ is related to the derivative of $\F(x)$ at zero.
Since $\frac{\d \F(x)}{\d x}|_{x=0}$ is the value of the PDF of $\ps$ at zero
and finite, it follows that $\kappa \geq 1$. We also have an upper bound on $\kappa$,
since $\F(x)$ is monotonically decreasing and its derivatives are bounded by
Assumptions A.1 to A.4. It is intuitive that
a good MAC protocol should spread out the transmissions, and not cluster them in
order to avoid collisions. If the
MAC protocol is such that the minimum distance between the transmitters does not
increase with decreasing density $\eta$, then the resulting transmitting set
will have high interference, and hence poor outage performance. The first theorem
makes the notion of {\em spreading}\footnote{This is different from the  MAC induced clustering in CDMA systems
\cite{yang2007inducing,haenggi-com-2007}, where the MAC protocol is optimized
for a fixed density of transmissions.} transmissions as $\eta\rightarrow 0$ more
precise.
\begin{theorem}
  \label{thm:conv1}
If the transmitting set $\Phi_t$ of density $\eta$ is such that
 \[B.1: \quad  \lim_{\eta\rightarrow 0} \sup_{x \in \R^2}\mathcal{K}_\eta(S_1+x)
< \infty,\]
	where $S_1=[0,1]^2$,
then $\kappa \geq 1$.
  \label{thm:kappa_bounds}
\end{theorem}
\begin{IEEEproof}
  See Appendix \ref{app:proof12}.
\end{IEEEproof}
 Let
    $C_\eta = \sup_{x\in\R^2}\mathcal{K}_\eta(x+S_1)$ where $S_1=[0,1]^2 $.
    Then from a similar	  argument as in the proof, it is easy to observe that
	  \[\EP[\Phi_t(B(o,b))] = \lambda\eta \mathcal{K}_\eta(B(o,b)) <
\lambda\eta \ceil{\pi
	  b^2}C_\eta.\]  Condition $A.1$ states that $\lim_{\eta \rightarrow 0}
C_\eta<\infty$, which implies
	  \[\EP[\Phi_t(B(o,\eta^{-a})] \rightarrow 0\qquad\text{for }a<1/2\,.\]
	So  $A.1$ implies that the average number of points in a
	ball of radius $b_\eta=\eta^{-a}$, $a<\frac{1}{2}$, goes to
	zero as the density tends to zero. This condition is violated when the
	average nearest-interferer distance remains constant with decreasing
density $\eta$.
See Figure \ref{fig:unreasonable} and \cite{ricardo-10} for examples that
violate Condition $B.1$. If $B.1$ is violated, $\p$ is not equal to $1$ when $\eta \rightarrow
0$.
\begin{theorem}
  \label{thm:upper1}
  If the transmit point process $\Phi_t$ is such that there exists a $R_1>0$
such that
  \begin{equation}
B.2:\quad\quad	\lim_{\eta \rightarrow 0} \eta K_\eta(R_1 \eta^{-1/2}) >0,
	\label{eq:C2}
  \end{equation}
 then $\kappa \leq  \frac{\nu\alpha}{2}$.
\end{theorem}
\begin{IEEEproof}
  See Appendix \ref{app:proof13}.
\end{IEEEproof}
 From Definition \ref{def:secorder}, $ \eta K_\eta(R_1
\eta^{-1/2})$ is equal to the average number of
points in a ball of radius $R_1 \eta^{-1/2}$ and hence Condition $A.2$
	requires the number of points inside a ball of radius
	$R_1\eta^{-1/2}$ to be greater than zero.
 Except for pathological
	cases, this condition is generally valid since the nearest-neighbor
distance scales like $\eta^{-1/2}$ when the point process has
	density $\eta$.
Essentially any MAC which schedules
	the nearest interferer only at an average distance 
$\Theta(\eta^{-1/2})$
satisfies these two Conditions $B.1$ and $B.2$, as most common MAC schemes do. {\em Henceforth we consider
only MAC schemes which satisfy Conditions  $B.1$ and $B.2$.}


We now investigate the scaling behavior of some common protocols, and the
interference exponent $\kappa$ and $\gamma$ values they achieve.  The average interference is
\begin{align}
  \EP[\I]&=\EP \sum_{\z \in \Phi_t}\h{\z r(o)}\l(\z-r(o)))\nonumber\\
  &\stackrel{(a)}{=} \E[\h{}]\eta^{-1} \int_{\R^2} \rho^{(2)}_\eta(x)
  \l(x-r(o))\d x,
  \label{eq:avg_inter}
  \end{align}
\chr{  where $(a)$ follows from the definition of the second-order product density. } The analysis of the scaling behavior of $\p$
depends on whether the average interference is finite or not.  The success probability is
$\EP\F(\T \I/\l(r))$, and if $\I$ is finite almost surely, then the Taylor
series expansion of $\F(x)$ can be used to obtain the asymptotics. On the other
hand, when $\EP[\I]=\infty$, then the Taylor series expansion cannot be used. For
example  consider an ALOHA Poisson network with a path loss model
$\l(x)=\|x\|^{-\alpha}$ and Rayleigh fading. As $\rho_\eta^{(2)}(x)
=\eta^{2}$ for a PPP,  the average interference from \eqref{eq:avg_inter} is
\[\EP[\I]= \eta\E[\h{}] \int_{\R^2} \|x\|^{-\alpha}\d x =\infty.\]
Hence, the Taylor series expansion of $\F(x)$ cannot be used. But nevertheless, in
this case \cite{bacelli-aloha}
\[\p=\exp(- \eta R^2 \T^{2/\alpha} C(\alpha)) \sim 1- \eta R^2 \T^{2/\alpha} C(\alpha), \quad \eta \to 0,\]
where $C(\alpha)= \pi \Gamma(1-2/\alpha)\Gamma(1+2/\alpha)$, which shows that $\kappa=1$.
In the next two sections, we analyze
the outage probability when the transmitting set $\Phi_t$ is selected from $\Phi$
using ALOHA and  CSMA, respectively. As we shall see, these two
MAC protocols achieve the extremes of the interference scaling exponent, with
ALOHA resulting in $\kappa=1$ and  CSMA  in $\kappa=\alpha\nu/2$.

\section{ALOHA  Networks}
\label{sec:ALOHA}
The completely decentralized nature of the ALOHA protocol makes it appealing for
a wireless network and particularly suitable for an ad hoc network. The
independent scheduling of the ALOHA process also makes its analysis much
easier than other plausible MAC protocols like CSMA that introduce  correlations between the transmit node locations.
ALOHA is a natural model when the underlying
node distribution is a PPP, since independently thinning a PPP results in a PPP
of lower density. Not surprisingly, this combination of PPP with the ALOHA MAC
protocol has been extensively analyzed, and
closed-form expressions for outage have been obtained for both SISO and MIMO systems \cite{bacelli-aloha,hunter2008transmission}.

\subsection{Asymptotic Success Probability}
An ALOHA network consists of a stationary point process $\Phi$ (need not be a PPP) of unit
density, where each node decides to transmit
independently of every other node with probability $\eta$. The resulting
transmitter set $\Phi_t$ is a stationary point process of density $\eta$,
and the interference is
\[\I = \sum_{\x \in \Phi} \h{\x r(o)}\l(\x-r(o))\i(\x \in \Phi_t),  \]
where $\i(\x \in \Phi_t)=1$ when the node $\x$ is selected by the ALOHA MAC
protocol to transmit.

First, as an example, consider a point process $\Phi=\{\x_1, \x_2\}$  composed of only
two points (which is a non-stationary network but suitable to illustrate our key
idea). Let
$ Y_\x=\h{\x r(o)}\l(\x -r(o)) $, $\forall \x \in \Phi$. After thinning $\Phi$
with ALOHA, the interference is
\[\I = Y_{1}\i(\x_1 \in \Phi_t) + Y_{2}\i(\x_2 \in \Phi_t).\]
Then,
{\em conditioned on $Y_1$ and $Y_2$},
by averaging $\F(\I)$ with respect to the ALOHA indicator random variables,
\begin{align*}\E \F(\I) = & (1-\eta)^{2}\F(0) +
\eta(1-\eta)\left(\F(Y_1)+\F(Y_2)\right)
+\eta^{2}\F(Y_1+Y_2).\end{align*}
Since $\F(0)=1$,
\[\E \F(\I)= 1-\eta\sum_{i=1}^2 \left(1-\F(Y_i)\right) +o(\eta), \quad \eta
\rightarrow 0.\]
Then averaging with respect to $Y_i$ (\ie, fading and point process $\Phi$), we
obtain that $\kappa=1$ and $\gamma = \EP \sum_{i=1}^2 (1-\F(Y_i))$. Theorem
\ref{thm:aloha_out} formalizes the above idea and shows that the interference scaling exponent $\kappa$ is
always $1$ for ALOHA.
\begin{theorem}
  \label{thm:aloha_out}
 When ALOHA is used as the MAC protocol, and a node from  a stationary point process $\Phi$  is allowed to transmit
  with probability $\eta$,  the success probability is
  \[\p\sim 1-\eta \gamma_{\mathtt{ALOHA}}, \quad \eta \rightarrow 0,\]
  where
  \[\gamma_{\mathtt{ALOHA}} = \int_{\R^2}
  \left[1-\E_{\h{}}\F\left(\h{}\frac{\T
  \l(x-r(o))}{\l(R)}\right)\right]\rho^{(2)}(x)\d x,\]
and  $\rho^{(2)}(x)$ is the second-order product density of the initial point process  $\Phi$.
  \label{thm:aloha}
\end{theorem}
\begin{IEEEproof}
  See Appendix \ref{app:aloha}.
\end{IEEEproof}
The above theorem indicates that $\kappa=1$ irrespective of the point process
$\Phi$ and $\F(x)$. The initial point process $\Phi$ affects the
asymptotic probability only by its second-order product density. It is also
evident that the channel diversity (by using multiple antennas) does not change the exponent $\kappa$,  but
only affects the pre-constant $\gamma_{\mathtt{ALOHA}}$.

{\em Intuition:} An independently thinned stationary point process
converges weakly to a PPP  as the density
  tends to zero \cite{verejones}. Hence, as $\eta\rightarrow 0$, ALOHA results in a point process
that closely resembles a PPP. The result follows since
$\kappa=1$ for a PPP. From Theorem \ref{thm:aloha},
\begin{align}
1-\p &\sim \eta\int_{\R^2}
  \left[1-\E_{\h{}}\F\left(\h{}\frac{\T
  \l(x-r(o))}{\l(R)}\right)\right]\rho^{(2)}(x)\d x\nonumber\\
  &= \EP \sum_{\x \in \Phi_t} 1-\E_{\h{}}\F\left(\h{}\frac{\T
  \l(\x-r(o))}{\l(R)}\right).
  \label{eq:interp}
\end{align}
Observe that $1-\E_{\h{}}\F\left(\h{}\frac{\T   \l(x-r(o))}{\l(R)}\right)$ is
the outage caused at the receiver $r(o)$ by a single interferer located at $x$. Hence, in the
low-density regime, the overall outage is equal to the sum of the outages
caused by each interferer.
To evaluate the transmission capacity,  the following lemma, which addresses
the monotonicity of $\p$, is required.
\begin{lemma}
  \label{lem:aloha_mon}
  Let $\p(\eta)$ denote the success probability when the ALOHA parameter is
  $\eta$. Then for any stationary point process $\Phi$,
  \[\eta_1\geq \eta_2 \Rightarrow \p(\eta_1)\leq \p(\eta_2).\]
\end{lemma}
\begin{IEEEproof}
  The proof follows from a standard uniform coupling argument
  \cite{bollobás2006percolation} and is omitted.
	  \end{IEEEproof}

In an ALOHA network, from the definition of $\gamma_{\mathtt{ALOHA}}$, we observe that if
$\rho^{(2)}_1(x) < \rho^{(2)}_2(x)$, $\forall x$  then $\gamma_1$ is smaller than $\gamma_2$.
So starting with a point process with a guaranteed minimum distance (such as the Matern
hard-core process) results in a better outage performance than a Poisson
network. For a Poisson cluster process of unit density $\rho^{(2)}(x) > 1$, and hence the
$\gamma$ for a cluster process is higher than for a PPP, resulting in a worse outage
performance.

\subsection{Example: Poisson ALOHA Networks}
We now provide a few examples and evaluate $\gamma_{\mathtt{ALOHA}}$ when the initial point process $\Phi$ is a PPP. When $\Phi$ is a Poisson point
process of unit density, the second-order product density is equal to
$\rho^{(2)}(x) = 1$, and the constant $\gamma_{\mathtt{ALOHA}}$ can be evaluated for different
$\F(x)$ using Theorem \ref{thm:aloha}.
\subsubsection{Nakagami-$m$ fading} When the fading is Nakagami-m distributed, $\F(x)$ is
the CCDF of the gamma
distribution with m-degrees of freedom,   given by
\[\F(x) = \frac{\Gamma(m,mx)}{\Gamma(m)}.\]
When $\l(x)=\|x\|^{-\alpha}$,
\begin{equation}
  \gamma_{\mathtt{ALOHA}}(m) =\frac{\pi\T^{2/\alpha}R^2\Gamma\left(m-2/\alpha\right)\Gamma\left(m+2/\alpha\right)}{
  \Gamma(m)^2}, \quad m\geq 1.
  \label{eq:nak}
\end{equation}
This constant has also been calculated  in
\cite{hunter2008transmission}. We observe that $\gamma_{\mathtt{ALOHA}}(m)$ decreases to $\pi$ as
$m\rightarrow \infty$, which shows that the outage performance improves as the
amount of fading decreases.

\subsubsection{Receiver beamforming}
When every transmitter uses a single antenna for transmission,
each receiver is equipped with $M_r$ receive antennas, and receive beamforming
is used, the
resulting  $\ps$ is  $\chi^2$
distributed with $2M_r$ degrees of freedom, and
\[\F(x)=\frac{\Gamma(M_r,x)}{\Gamma(M_r)}.\] For $\l(x)=\|x\|^{-\alpha}$, and
Rayleigh fading
\begin{equation}
  \gamma_{\mathtt{ALOHA}}({M_r})= \frac{\pi\T^{2/\alpha}R^2\Gamma(M_r-2/\alpha)\Gamma(1+2/\alpha)}{\Gamma(M_r)},
  \quad M_r\geq 1.
  \label{eq:mimo}
\end{equation}
Since $\gamma_{\mathtt{ALOHA}}({M_r})\approx \pi\T^{2/\alpha}R^2
\Gamma(1+2/\alpha)M_r^{-2/\alpha}$,  the
gain of using multiple receive antennas is $\Theta(M_r^{-2/\alpha})$, resulting
in a lower outage probability for large $M_r$.

 \section{CSMA Networks}
 \label{sec:CSMA}
We now consider MAC protocols which result in transmitter point processes
$\Phi_t$ that have a
minimum distance between the transmitters. In this case $\E[\I]<\infty$, and we use the Taylor series
expansion of $\F(x)$ to obtain the scaling of the success probability. For the sake of illustration,
consider only one interferer at a distance $\eta^{-1}$ from $r(o)$ and let
$\F(x) =1-x^{2} +x^{3} +o(x^3)$,  $x \rightarrow 0$. Neglecting the fading of the interferers,  observe that
\[\F(\I) = 1-\eta^{\alpha}+\eta^{3\alpha/2} = 1-\eta^{\alpha}
+o(\eta^{\alpha}),\]
which shows that $\kappa =\alpha$. In this section, we formalize this procedure
to obtain the asymptotic behavior. We use CSMA as a representative MAC protocol
and obtain the asymptotic outage probability. For other MAC protocols that impose a
minimum distance between transmitters, this proof technique can be easily adapted.

\subsection{ CSMA Model}
\begin{figure}[ht]
  \begin{center}
	 \includegraphics[width=0.45\columnwidth]{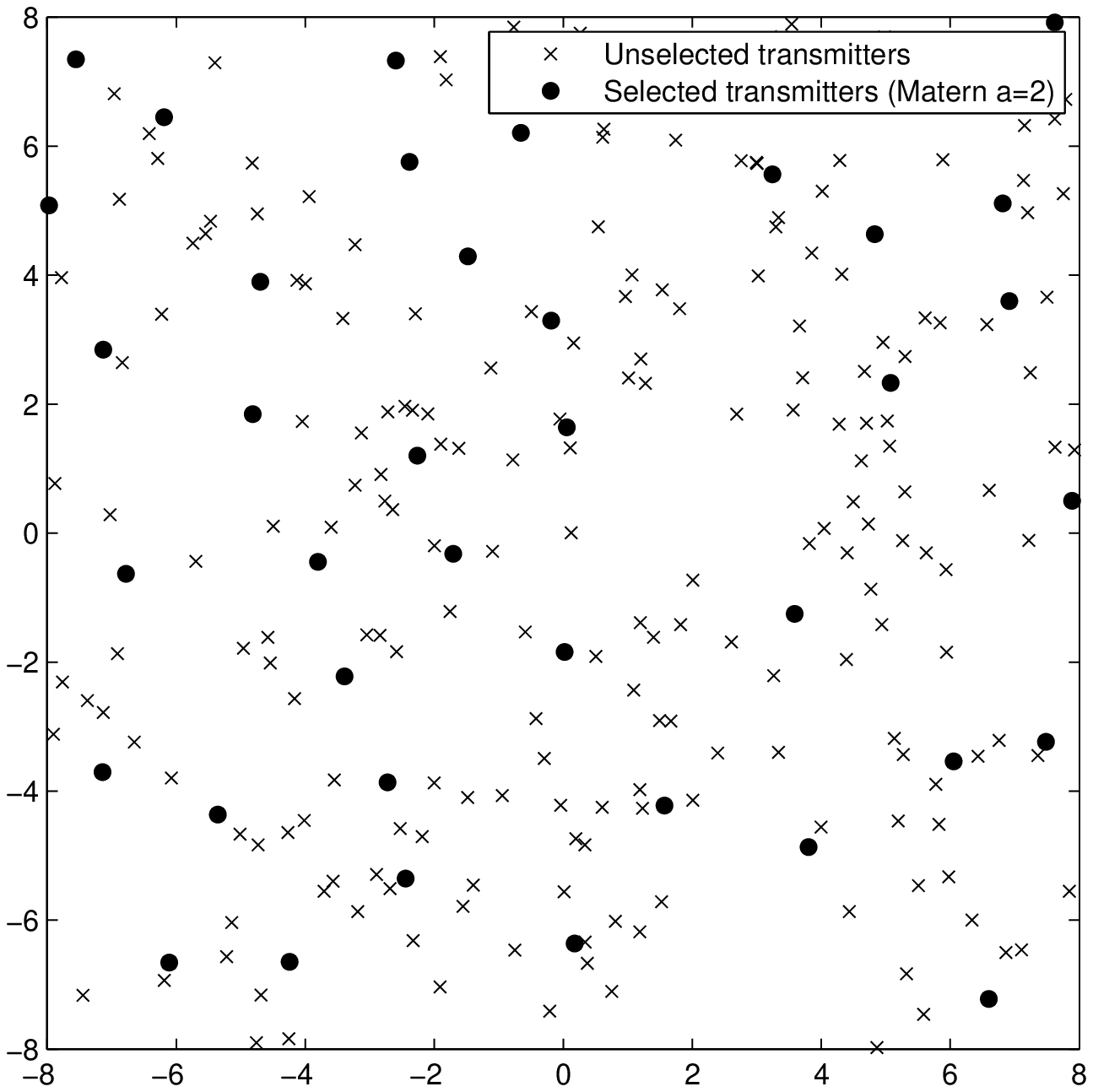}\includegraphics[width=0.45\columnwidth]{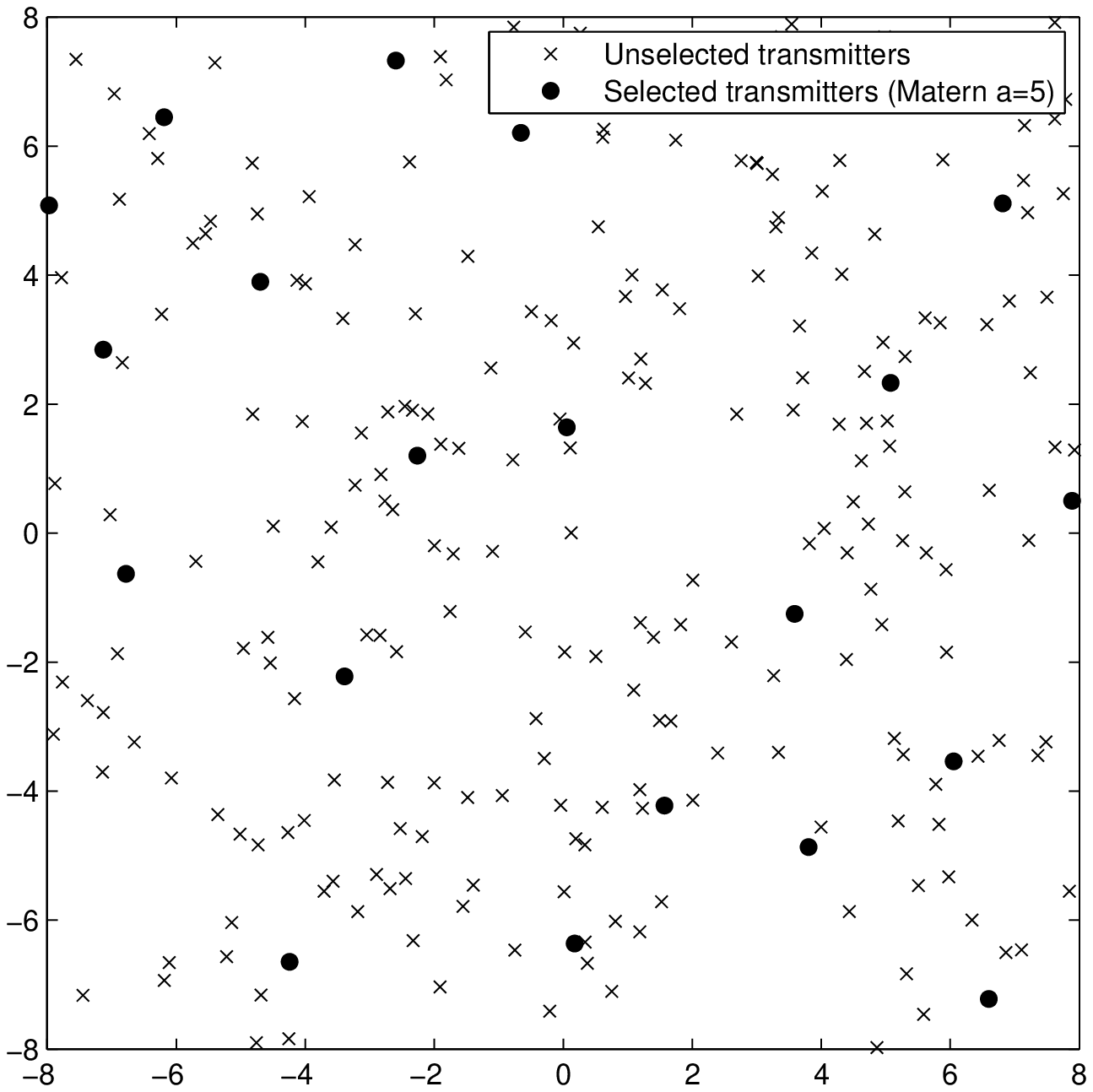}
  \end{center}
  \caption{ Left: Matern hard-core process with $\a=2$. Right: Matern hard-core
  process with $\a=5$.  Observe that the  minimum distance between nodes increases
  with $\a$. }
  \label{fig:matern}
\end{figure}

Although the spatial  distribution  of the transmitters that concurrently
 transmit in  a CSMA network is difficult to determine, the transmitting set
can be closely approximated by a Matern hard-core process
\cite{stoyan,BaccelliNOW}.
Start with a  Poisson point process  $\Phi$ of
unit density. To each node $\x \in \Phi$,  associate a mark $m_\x$, a uniform random
variable in $[0,1]$. The contention neighborhood of a node $\x$ is the set of
nodes which result in an interference power of  at least    $\pp$ at $\x$, \ie,
\[\bar{\calN}(\x) = \{\y \in \Phi: \l(\y -\x) > \pp \}.\]
Observe that we are not including fading in the selection of the contention
neighborhood. An alternative model has been proposed in \cite{BaccelliNOW} which
incorporates the fading between the nodes for the selection of the contention
set. A node $\x \in \Phi$ belongs to the final CSMA transmitting set if
\[ m_\x < m_\y,\quad \forall \y\in \bar{\calN}(\x). \]
Define $\a = \l^{-1}(\pp)$. See Fig. \ref{fig:matern} for an illustration of
Matern hard-core process.
 The average number of nodes in the contention neighborhood of $\x \in \Phi$
does not depend on the location $\x$ by the stationarity of $\Phi$ and is equal
to \cite{BaccelliNOW}
\[\calN = \E[|\bar{\mathcal{N}}(x)|]= \pi \a^2.\]
The density of the CSMA Matern process $\Phi_t$ is
\[\eta =  \left[\frac{1-\exp(-{\calN})}{{\calN}} \right].\]
Let $\rho^{(k)}_\a(x_1,\dots x_{k-1})$ denote the $k$-th order product density
of the Matern hard-core point process $\Phi_t$.
Let $\BB(x_1,\dots,x_n)$ denote the volume of the intersection
of discs of radius $\a$ centered around $x_i$ with the convention $\BB(x_i)=\pi
\a^2$. Let $x_k$ denote the origin $o$, and for any indexing set $J=\{a_1,\dots,a_{|J|}\}$, and  let
$\BB(J)= \BB(x_{a_1}\dots x_{a_{|J|}})$. The following lemma characterizes the
product density and its scaling behavior.
\begin{lemma}
  The $k$-th order product density satisfies
  \[\rho^{(k)}_\a(\a x_1,\dots \a x_{k-1})\sim \a^{-2k}\tilde{\rho}^{(k)}(x_1,\dots
  x_k),\quad \a \rightarrow \infty, \]
  where
 \begin{align}
   \tilde{\rho}^{(k)}( x_1,\dots, x_{k-1})
& = k!\int_{\R^k}  \exp\left(-\sum_{J\subset\{1,\dots,k \}} (-1)^{|J|+1}m_{\min
 J} \BBB( J)\right)\nonumber\\
  &\cdot\i(0\leq m_1\leq\dots\leq m_k<\infty)\d m_1\dots\d m_k.
  \label{eq:prod_inf}
\end{align}
for $\|x_i\| >1$,  $ 1\leq i\leq k-1$, and $0$ otherwise.
  \label{lem:prod}
\end{lemma}
\begin{IEEEproof}
  See Appendix \ref{app:prod}
\end{IEEEproof}
\subsection{Asymptotic Success Probability}
We now present the main result that characterizes the outage behavior of a CSMA
system at low densities. Before that, we require the following notation.
 Let
  $\mathcal{P}(n)$ denote the set of partitions of $n$, \ie, the set of integers
  $(p_1,
  p_2\dots,p_k)$, such that $\sum p_i = n$ and $1\leq p_i \leq n$. Tuples with
  different ordering are not distinguished.
 For example
$\mathcal{P}(2) =\{(2),(1,1)\}$, and $\mathcal{P}(4)
=\{(4),(1,1,1,1),(1,2,1),(3,1),(2,2) \}$.
\begin{theorem}
  \label{thm:matern}
  When the set of transmitters forms a Matern CSMA process of density $\eta$ and
  the Conditions A.1, A.2 and A.4 are satisfied, then
  \begin{equation}
	\p \sim 1- c_0 \pi \left(\frac{\T}{\l(R)}\right)^{\nu} \eta^{\alpha\nu/2}A_I,\quad
  \eta \rightarrow 0,
  \label{eq:csma12}
\end{equation}
  where $A_I$ is given by
  \begin{equation}
	\sum_{(p_1,\dots,p_k)\in \mathcal{P}(\nu)}
	\int_{\R^{2k}}   \tilde{\rho}^{(k+1)}(x_1,\dots,x_k)
\prod_{i=1}^{k} \|x_i\|^{-\alpha p_i} \E[\h{}^{p_i}]\d
  x_i,
  \label{eq:csma11}
\end{equation}
where $\tilde{\rho}^{ (k+1) }(x_1,\dots,x_k)$, $1\leq k\leq \nu$, is given in Lemma \ref{lem:prod}.
\end{theorem}
\begin{IEEEproof}
  The proof is presented in Appendix \ref{app:csma}.
\end{IEEEproof}
From \eqref{eq:csma11}, we observe that the outage probability $1-\p$ at low
density can be decomposed into three parts:
\begin{enumerate}
  \item $c_o$ stems from the source-destination distribution CCDF $\F(x)$.
  \item \chr{The interference-scaling exponent $\kappa=\alpha\nu/2$ depends on the CCDF $\F(x)$.}
  \item $A_I$ depends on the point process and the fading between the destination
	$r(o)$ and the interferers, specifically on the $\nu$  moments of fading
	$\E[\h{}^p]$, $1\leq p\leq \nu$.
\end{enumerate}
So from a system design perspective, one has to reduce $c_0$  and $A_I$
for a better outage performance.
For a given $\nu$, reducing $A_I$ would result from reducing $\E[\h{}^p]$, $1\leq p\leq \nu$.
 The following Lemma proves the monotonicity of success probability in a CSMA
 network with respect to the  exclusion radius.
\begin{lemma}
  \label{lem:matern_mon}
Let $\p(\a)$ denote the probability of success in a Matern hard-core process when
the exclusion radius is $\a$.  For a given initial Poisson point process $\Phi$,
applying the Matern hard-core process with radii $\a_1$ and $\a_2$, such that
$\a_1>\a_2$, implies
\[\p(\a_1)>\p(\a_2).\]
\end{lemma}
\begin{IEEEproof}
  Fix the marks of $\Phi$ uniformly distributed in $[0,1]$. It is then obvious
  that the Matern process generated with $\a_1$ as the exclusion radius is a subset of the Matern process
  generated with an exclusion radius $\a_2$.
\end{IEEEproof}

  The proof of  Theorem \ref{thm:matern} can be easily adapted  to any MAC protocol that  results in a transmitter set where
   the product densities satisfy
	\begin{equation}
	  \int_{\R^{2k}}
	  \rho^{(k+1)}_\eta(x_1,\dots,x_k)\prod_{i=1}^k\|x_i\|^{-\alpha p_i}\d
	  x_i < C_4^{k}\eta^{ \frac{1}{2}\alpha\sum_{i=1}^k p_i }
	  \label{eq:cc22}
	\end{equation}
	for some $C_4 >0$, $ \forall k \geq 1$, and  for all integers $ p_i\geq 1$.
We require bounds on all the product densities in \eqref{eq:cc22} because we are
considering arbitrary $\F(x)$. If the function $\F(x)$ has bounds of
the type
\[ 1-c_0 x^{\nu} \leq \F(x) \leq 1-c_0x^{\nu}+ c_1x^{\nu+1}, \]
then the asymptotic theorem can be proved by just considering $\nu+2$ product
densities. In this case it suffices to prove that $\EP[\I^{\nu+1}] =
o(\EP[\I^{\nu}])$ as $\eta\rightarrow 0$.
For example, when $\F(x)=\exp(-x)$  we have
\[1-x\leq \F(x)\leq 1-x+x^2, \quad  x>0,\] and hence  the second and the  third order
product densities are sufficient to evaluate the asymptotics.

\section{Effect of Noise}
\label{sec:noise}
In the previous sections, we have neglected noise in the asymptotic outage analysis. We now discuss how noise changes the value of $\gamma$ and $\kappa$.    With thermal noise, the success probability is 
\[\p=\PP\left(\frac{\ps \l(R)}{\I+\frac{\sigma^2}{P} }>\T\right),\]
where $\sigma^2$ is the noise power, and $P$ is the common transmit power.  Using the CCDF of $\ps$ we obtain
\[\p=\EP\F\left(\frac{\T\I}{\l(R)}+\W \right),\]
where $\W =\frac{\sigma^2\T}{P\l(R)} $.
So when the density of transmitters $\eta\to 0$,  we no longer have $\pn$ equal to $1$, where
\[\pn \triangleq \lim_{\eta \to 0} \p  = \F\left(\W \right).\] 
Hence for small  $\eta$, the success probability can be  represented as
\begin{equation}
\p \sim  \pn-\gamma \eta^\kappa,\quad \eta \to 0.
\label{eq:noi}
\end{equation}
Consider the Taylor series expansion of $\F(x)$ around $\W$ 
  \begin{equation}
   \F(x+\W)= 1-c_o(\W) x^{\nu}+\sum_{k=1}^{\infty}\frac{c_k(\W)}{k!}x^{k+\nu}, \quad
  c_o(\W)\neq 0.
  \label{eq:tay}
  \end{equation}
  Since $1-\F(y)$ is the CDF, the derivative of $1-\F(y)$  corresponds to the PDF. For any unimodal distribution, the PDF is non-zero at $y\neq 0$, and hence it follows that  $\nu =1$ for $\W\neq 0$.  The following observations can be made concerning the spatial contention parameter $\gamma$ and the interference scaling exponent $\kappa$ when noise is present. 
  \begin{enumerate}
  \item Since $\nu=1$, the interference scaling exponent $1\leq \kappa \leq \alpha/2$.  So unlike the noiseless case, the upper bound on $\kappa$ depends only on the path loss exponent.
  
  \item    With ALOHA, the interference scaling exponent $\kappa=1$, while the spatial contention parameter is 
  \begin{equation}
  \gamma=\int_{\R^2}
  \left[\F(\W)-\E_{\h{}}\F\left(\h{}\frac{\T
  \l(x-r(o))}{\l(R)}+\W\right)\right]\rho^{(2)}(x)\d x.
  \label{eq:aloha_noise}
  \end{equation}
  This can be obtained using techniques similar to the proof of  Theorem  \ref{thm:aloha}.
  \item When CSMA is used as the MAC protocol, $\kappa =\alpha/2$, and the spatial contention parameter  is
  \begin{equation}
  \gamma = \frac{\T c_0(\W) \pi \E[\h{}]}{\l(R)}\int_{\R^{2}}   \tilde{\rho}^{(2)}(x)
\|x\|^{-\alpha } \d x,
\label{eq:csma_no}
\end{equation}
 where $ \tilde{\rho}^{(2)}(x)$ is given by \eqref{eq:prod_inf}. This result can be obtained from the  proof of Theorem \ref{thm:matern}, by setting $\nu=1$.
    \end{enumerate}
 Analyzing outage with noise is particularly simple when $\F(x)=\exp(-x)$, \ie, Rayleigh fading.  
   The success probability is 
   \[\p = \exp\left(-\frac{\T \I}{\l(R)}\right)\exp(-\W).\]
Observe that  the success probability factorizes into two terms, one that depends only on interference and the other only on noise \cite{bacelli-aloha}. Using this factorization,  and  Theorem \ref{thm:aloha_out} it follows that
\[\gamma=\exp(-\W)\int_{\R^2}
  \left[1-\E_{\h{}}\F\left(\h{}\frac{\T
  \l(x-r(o))}{\l(R)}\right)\right]\rho^{(2)}(x)\d x,\]
  which agrees with \eqref{eq:aloha_noise}. Similarly it can be verified that $c_0(\W)=\exp(-\W)$ in  \eqref{eq:csma_no}.

\section{Transmission Capacity}
\label{sec:TC}
The transmission capacity (TC) metric captures the global effect of interference
on the network throughput. It depends on the spatial distribution of
transmitters, includes the MAC protocol, the
physical layer technologies used, the path loss function, and the target $\sir$
and outage probability. In short, the TC captures many of the most
important features of a wireless network.
%
Since for many MAC protocols, spatial distribution of nodes, and fading
distributions, the outage probability itself is not known, finding the
exact TC is not feasible.  Therefore, we analyze the TC in the asymptotic regime when
$\epsilon \rightarrow 0$. More precisely, we try to find an easily computable
function $\TCL(\epsilon)$ such that
\begin{equation}
\TC(\epsilon) = \TCL(\epsilon)+o(\TCL(\epsilon)),\quad \epsilon \rightarrow 0.
\label{eq:999}
\end{equation}
\chr{In the previous sections, we have obtained an asymptotically tight approximation of the success probability of the form 
\[\p= 1-\gamma \eta^\kappa, \quad \eta \to 0.\]
It is intuitive to expect that  the TC can be obtained  by inverting the asymptotic of $\p$, to solve for the density $\eta$  corresponding to $\p =1-\epsilon$.  The following lemma formalizes this asymptotic inversion and will be used  to characterize the TC.}
\begin{lemma}
  \label{lem:approx}
  Let $f(x)$ and $f_1(x)$ be  continuous, positive, strictly increasing
  functions with $f(0)=f_1(0)=0$. \chr{Furthermore let $f_1(x)$ be locally convex at $0$},  and
  \begin{equation}
	f(x) = f_1(x) +o(f_1(x)), \quad x\rightarrow 0.
	\label{eq:conv}
  \end{equation}
  Define $f^*(\epsilon) = \sup\{x: f(x) <\epsilon \}$, and $f_1^*(\epsilon)$
  similarly. Then
  \[f^*(\epsilon) = f_1^*(\epsilon) +o( f_1^*(\epsilon)),\quad \epsilon
  \rightarrow 0.\]
\end{lemma}
\begin{IEEEproof}
See Appendix \ref{app:asym}.
\end{IEEEproof}
 We will now use this Lemma  to obtain the asymptotic TC in ALOHA and CSMA networks.

\subsection{TC of ALOHA  Networks}

For an ALOHA network   $\kappa =1$, \chr{and hence the approximation $\gamma\eta$ is linear}. The monotonicity of $\p$ for ALOHA  follows from Lemma \ref{lem:aloha_mon}. Hence all the  necessary conditions of Lemma \ref{lem:approx} are satisfied in the case of ALOHA networks. Defining  $f(\eta) = 1-\p$, and using the asymptotic approximation of $\p$, we can obtain the TC for ALOHA networks from Lemma \ref{lem:approx}.

\begin{theorem}
 When ALOHA is used as the MAC protocol on a node set distributed as a
 stationary point process  $\Phi$ with second-order product density $\rho^{(2)}(x)$,
  \begin{equation}
	\TC(\epsilon) = \frac{\epsilon}{ \int_{\R^2}\left[1-\E\F\left(\h{}\frac{\T \l(x-r(o))}{\l(R)}\right)\right]\rho^{(2)}(x)\d x } +o(\epsilon)
  \end{equation}
  \label{thm:TC_ALOHA}
\end{theorem}
\begin{IEEEproof}
  Follows from Lemmas \ref{lem:aloha_mon} and \ref{lem:approx} and Theorem \ref{thm:aloha_out}.
\end{IEEEproof}
\chr{ Using Theorem \ref{thm:TC_ALOHA},   the asymptotic TC of PPP ALOHA networks, when $\ps$  and $\h{}$ are Nakagami-$m$ distributed or $\chi^2$ distributed,
can be obtained in closed form.}
   \begin{cor} [Nakagami-m fading]
When $\ps$ and $\h{}$ are Nakagami-m distributed, from Theorem  \ref{thm:TC_ALOHA}
the transmission capacity is
\[\TC(\epsilon) = \frac{\epsilon  }{ \gamma_{\mathtt{ALOHA}}(m)} +o(\epsilon),\]
where $ \gamma_{\mathtt{ALOHA}}(m)$ is given in \eqref{eq:nak}. 
\end{cor}

\begin{cor}[SIMO system with beamforming] When receiver beamforming with $M_r$
antennas is used, the TC is
\[\TC(\epsilon) = \frac{\epsilon }{ \gamma_{\mathtt{ALOHA}}({M_r})} +o(\epsilon),\]
where $ \gamma_{\mathtt{ALOHA}}({M_r})$ is given in \eqref{eq:mimo}.
\end{cor}
\chr{This asymptotic TC  obtained in both these cases can be verified \ch{by comparing them} with the results in  \cite{hunter2008transmission}, wherein the TC was obtained using different techniques.}

\subsection{TC of CSMA Networks}
Similar to ALOHA networks, the following theorem characterizes the asymptotic TC of CSMA networks.
\begin{theorem}
When the transmitting set of nodes results from the  CSMA MAC protocol described
in Section \ref{sec:CSMA}, the TC is
\begin{equation}
  \TC(\epsilon) = \frac{\epsilon^{\frac{2}{\alpha\nu}}}{\left(\pi c_oA_I
  \left(\frac{\T}{\l(R)}\right)^{\nu}\right)^{\frac{2}{\alpha\nu}}} +
 o(\epsilon^{\frac{2}{\alpha\nu}}),
\end{equation}
where $A_I$ is defined in Theorem \ref{thm:matern}.
  \label{thm:TC_CSMA}
\end{theorem}
\begin{IEEEproof}
  Follows from Lemmas \ref{lem:matern_mon},   \ref{lem:approx}, and Theorem \ref{thm:matern}.
\end{IEEEproof}
We note that the TC achieved by the
CSMA protocol is $\Theta(\epsilon^{\frac{2}{\alpha\nu}})$ compared to
$\Theta(\epsilon)$ for ALOHA. We also observe that the TC of CSMA  changes
by an exponent  for MIMO compared to SISO, while such gain is not registered by
ALOHA. \chr{We have the following simplification in the case of Rayleigh fading.}
\begin{cor}
 For CSMA with Rayleigh fading, \ie,  $\F(x) =\exp(-x)$,
 \[\TC(\epsilon) =  \left(\frac{\epsilon}{\gamma_{\mathtt{CSMA}}}\right)^{2/\alpha} +o(\epsilon), \]
where
 \begin{equation}
\gamma_\mathtt{CSMA}=\frac{R^\alpha \theta\pi^{\alpha/2}2^{3-\alpha}}{\alpha-2}+4\theta R^\alpha\pi^{2}\int_{1/\sqrt{\pi}}^{2/\sqrt{\pi}}\frac{r^{1-\alpha}}{g(r)}\d r\,,
\label{gamma_hc}
\end{equation}
and \[
g(r)=2\pi-2\arccos\left(\frac{\sqrt{\pi}}{2}r\right)+\frac{r\sqrt{\pi}}{2}\sqrt{4-\pi r^{2}}.\]
\end{cor}

\subsection{TC of General Networks with Rayleigh Fading}
In this subsection, we consider a network with a general distribution of nodes,
but restrict the CCDF $\F(x)$ to be exponential, and obtain the TC.  The distribution of fading between the interferers and the tagged receiver $\h{}$ can be arbitrary. \chr{The results in this subsection are extensions of Theorem \ref{thm:TC_ALOHA} and Theorem \ref{thm:TC_CSMA}  to general node distributions with exponential distribution for $\ps$.}

Define
  \begin{align*}
  \Delta(x) &\triangleq 1- \mathcal{L}_{\h{}}\left(\frac{\T}{\l(R)}\l(x-r(o)) \right),\\
\m & \triangleq \eta^{-1}\int_{\R^2} \sp(x) \Delta(x)\d x,\\
\k& \triangleq\eta^{-1}\int_{\R^2\times \R^2} \tp(x,y) \Delta(x)\Delta(y)\d  x\d y,
  \end{align*}
where  $\mathcal{L}_{\h{}}(x)=\E[e^{-\h{}x}]$ is the Laplace transform of the
  random variable  $\h{}$.  \chr{Since $\ps$ is exponential,}
  \[\Delta(x) = 1-\P\left(\ps>\frac{\h{}\T\l(x-r(o))}{\l(R)} \right).\]
Hence,  $\Delta(x)$ can be interpreted as
 the outage caused  by a single interferer  located at $x$.
Obviously  $0\leq \Delta(x)\leq 1$, and
 when $\h{}$ is gamma distributed with unit mean (Nakagami-m fading),
 \[\Delta(x) = 1-\left(1+\frac{\T \l(x-r(o))}{m\l(R)}\right)^{-m}.\]
The expectation of $\sum_{\x \in \Phi_\eta} \Delta(\x)$ is given by
 $\m$, and $\k$ denotes the average of $\sum_{\x ,\y \in \Phi_\eta, \x \neq \y}
 \Delta(\x)  \Delta(\y)$.
We now provide bounds on the success probability $\p$ which we later show to be
asymptotically tight.
\begin{lemma}
When $\F(x)=\exp(-x)$,  the probability of success is bounded for all $0\leq
\eta \leq 1$,
  \begin{equation}
	1-\m \leq \p \leq \min\{1-\m+\frac{\k}{2}, \mathcal{G}[\exp(-\Delta(x))]\},
	\label{eq:bound1}
  \end{equation}
  where $\mathcal{G}$ is the probability generating functional of $\Phi_t$ with respect to its reduced Palm measure.
 \label{thm:bounds}
 \end{lemma}
\begin{IEEEproof}
 The success probability is
 \begin{eqnarray*}
   \p &=& \PP\left(\ps\geq \T\l(R)^{-1}\sum_{\x \in \Phi_t} \h{\x
   r(o)}\l(\x-r(o)) \right)\\
   &\stackrel{(a)}{=}& \EP\exp(-\T\l(R)^{-1}\sum_{\x \in \Phi_t} \h{\x
   r(o)}\l(\x-r(o)) ),
 \end{eqnarray*}
 where $(a)$ follows since $\ps$ is an exponential random variable. Taking the
 expectation with respect to $\h{\x r(o)}$ in the interference,
 we obtain
 \[\p = \E^{!o}\left[ \prod_{\x \in \Phi} 1-\Delta(\x) \right].\]
 We can observe that $0\leq \Delta(x) \leq 1$, and hence using the inequality
 \[ 1-\sum a_i \leq \prod 1-a_i \leq 1-\sum a_i
 +\frac{1}{2}\sum_{i\neq  j} a_i a_j,\] and the definition of the second- and third-order product densities, we obtain
 the lower bound $1-\mu$, and the upper bound $1-\mu +\kappa/2$.  The other upper
 bound can be obtained  using the inequality $1-\Delta(\x) \leq
 \exp(-\Delta(\x))$ and the definition of the probability generating functional.
\end{IEEEproof}
Define  
\[\TCU(\epsilon)  \triangleq  (1-\epsilon) \sup\{\eta:  \min\{1-\m+\k/2,
\mathcal{G}[\exp(-\Delta(x))]\} \geq 1-\epsilon\},\]
 and
 \[\TCL(\epsilon)\triangleq (1-\epsilon)\sup\{\eta: \m \leq \epsilon \}.\] 
It can be easily verified that 
\begin{equation}
  \TCL(\epsilon) \leq \TC(\epsilon) \leq \TCU(\epsilon).
  \label{eq:TCineq}
\end{equation}
We now present the main theorem of this subsection which characterizes the TC of general networks when $\ps\sim \exp(1)$.

  \begin{theorem}
	\label{thm:main}
   If the transmitter set $\Phi_t$ satisfies Condition B.1, 
   \begin{description}
\item[C.1]  $\lim_{\eta \rightarrow 0}
  \frac{\eta^{-1}\int_{\R^2}\rho^{(2)}_\eta(x)\Delta^{2}(x)\d x}{\k} >2, $
	 \item[C.2] $\k=o(\m)$,
   \end{description}
   as $\eta \rightarrow 0$,  the transmission capacity  when $\F(x)=\exp(-x)$ is 
	\[\TC(\epsilon) = \TCL(\epsilon)+o(\TCL(\epsilon)), \quad \epsilon
	\rightarrow 0.\]
	  \end{theorem}
  \begin{IEEEproof}
	See Appendix \ref{app:main}.
  \end{IEEEproof}

  {\em Remarks:}
  \begin{enumerate}
   \chr{
  \item Computing $\TCL(\epsilon)$ is much simpler than computing the actual TC, since $\rho^{(2)}(x)$  is known for many point processes. 
  \item In the case of ALOHA networks,  $\rho_\eta^{(2)}(x) = \eta^2 \rho^{(2)}(x)$, and hence 
  \[\TCL(\epsilon)= \frac{\epsilon (1-\epsilon) }{\int_{\R^2} \rho^{(2)}(x)\Delta(x)\d x},\] which matches Theorem \ref{thm:TC_ALOHA} with $\F(x)=\exp(-x)$.
    \item Conditions C.1 and C.2 are generally satisfied by most point processes. For example for a PPP of density $\eta$, the second-order product density  $\rho^{(2)}_\eta(z) =\eta^2$,  and hence
   \[\m = \eta\int_{\R^2} \Delta(x) \d x. \]
Also,  $\rho^{(3)}(x,y) = \eta^3$, and  so
\[\k= \eta^2\int_{\R^2\times \R^2	} \Delta(x)\Delta(y)\d x\d y =\mu_\eta^2 .\]
Therefore it follows that $\k =o(\m)$. Also
\[\lim_{\eta \rightarrow 0}
  \frac{\eta^{-1}\int_{\R^2}\rho^{(2)}_\eta(x)\Delta^{2}(x)\d x}{\k} = \lim_{\eta\to 0}\eta^{-1}\frac{\int_{\R^2} \Delta^2(x)\d x}{(\int_{\R^2}\Delta(x))^2}=\infty>2,\]
  verifying C.2. In fact, constructing  {{\em reasonable}} MAC schemes \cite{giacomelli-asymptotic} that lead to a transmitter process $\Phi_t$ for which Conditions C.1 and C.2 are violated is
  difficult\ch{---if not impossible.}}
  \end{enumerate}
  
\section{Conclusions}
 
 \chr{A new mathematical framework to analyze outage in networks with complex MAC protocols,   \ch{general} spatial node distributions, and  general fading  is introduced.} We \ch{define} two constants, \ch{the interference scaling exponent} $\kappa$ and \ch{the spatial contention} $\gamma$ such that the success probability $\p\sim 1-\gamma
\eta^{\kappa}$, $\eta \rightarrow 0$, where $\eta$ is
the density of transmitters. By the continuity of $\p$, this formulation
characterizes the outage probability for small $\eta$.
We also use this framework to analyze the transmission capacity. We
first show that the interference scaling exponent \chr{satisfies} $1\leq \kappa \leq \alpha
\nu/2$, where $\alpha$ is the path loss exponent and $\nu$ depends on the
distribution of the received power.  For ALOHA  we
show that $\kappa=1$, irrespective of the spatial node distribution we start
with or the channel diversity. We also provide the constant $\gamma$ in this
case. The other extreme of the interference scaling exponent, $\kappa
=\alpha\nu/2$, is obtained by the CSMA MAC protocol, for which we also provide the exact
value of $\gamma$. The constant $\gamma$ depends only on the product densities
of the transmitter point process. Using this
framework we also obtain the transmission capacity of ALOHA and CSMA networks
for any channel gain distribution. We show that for a CSMA network the TC scales
like $\Theta(\epsilon^{\frac{2}{\alpha \nu}})$, in contrast to an ALOHA network with a TC
of $\Theta(\epsilon)$. When the channel gains are Rayleigh distributed, we
obtain the asymptotic TC for any MAC protocol and spatial distribution of
nodes. The mathematical techniques introduced in this paper can be used to
analyze other relevant metrics for a low density of interferers.

\appendices
%
\section{Proof of Theorem \ref{thm:conv1}}
\label{app:proof12}
  It suffices to prove that $1-\p =o(\eta^{1-\epsilon})$ for any $\epsilon >0$.
  For notational convenience, without loss of generality we assume $\T/\l(R)
  =1$. The probability of success is equal to $\EP \F(\I)$ and hence
  \[\frac{1-\p}{\eta}= \frac{\EP[1-\F(\I)]}{\eta}.\]
  Since $\F(x)$ is a Lipschitz function and also a CCDF, we have
  \[\F(x)-\F(x+y)\leq \min\{1,C y\}, \quad y \geq 0.\]
  Hence we obtain
  \begin{equation}
  \frac{\EP[1-\F(\I)]}{\eta} \leq
  \frac{\EP[\min\{1,C\I\}]}{\eta}\leq  \frac{\max\{1,C\}\EP[ \min\{1,\I\}]}{\eta}.
  \label{eq:lips}
\end{equation}
        Define $b=\l^{-1}(1)$.
\chr{         Define $\Phi_t^{\mathrm{near}}=\{\x \in \Phi_t: \l(\x-r(o))>1 \}$, and  $\Phi_t^{\mathrm{far}}=\{\x \in \Phi_t: \l(\x-r(o))<1 \}$}. 
Then it follows that
\chr{
 \begin{align}
   \min\{1,\I\} &\leq  \min\left\{ 1,\sum_{x\in\Phi_t^{\mathrm{near}}} \h{\x
r(o)}\l(\x-r(o))+ \sum_{\x \in\Phi_t^{\mathrm{far}}} \h{\x
r(o)}\l(\x-r(o))\right\},\nonumber\\
 &\stackrel{(a)}{\leq}\sum_{x\in\Phi_t^{\mathrm{near}} }\min\left\{ 1, \h{\x
r(o)}\l(\x-r(o))\right\} + \sum_{\x \in\Phi_t^{\mathrm{far}}} \h{\x
r(o)}\l(\x-r(o)),\nonumber\\
   &\leq \underbrace{\Phi_t( B(r(o),b))}_{\I_1}+ \underbrace{\sum_{\x \in\Phi_t^{\mathrm{far}} } \h{\x
r(o)}\l(\x-r(o))}_{\I_2}.
\label{eq:sam}
\end{align}
To obtain $(a)$, we have used the fact that $\min\{1,x+y\}\leq \min\{1,x\}+\min\{1,y\}$. Also recall that $\Phi_t( B(r(o),b))$ denotes the number of of $\Phi_t$ in the set $B(r(o),b)$.}
  Using the definition of the second-order product density, we obtain
  \begin{align*}
	\eta^{-1}\EP[\I_2] &= \eta^{-1}\EP\sum_{\x \in\Phi_t^{\mathrm{far}} }
\h{\x r(o)} \l(\x-r(o))\\
	&=\eta^{-1}\E[\h{}]\EP\sum_{\x \in\Phi_t^{\mathrm{far}}} \l(\x-r(o))\\
	&\stackrel{(a)}{=}\E[\h{}]\eta^{-2}\int_{\l(x-r(o))<1} \rho^{(2)}(x) \l(x-r(o)) \d x,
  \end{align*}
  where $(a)$ follows from the definition of $\sp(x)$.
Define $\tilde{\l}(x) = \l(x)$, when $\l(x-r(o))<1$ and $\tilde{\l}(x)=1$
otherwise. We then  have
\begin{align*}
  &\eta^{-2}\int_{\l(x-r(o))<C}\rho^{(2)}(x)\l(x-r(o))\d x\\
&  \leq \eta^{-2}\sum_{(k,j)\in
\Z^2}\int_{S_{kj}}\rho^{(2)}(x)\tilde{\l}(x-r(o))\d x,\end{align*}
where $S_{kj} =[k, k+1]\times [j, j+1]$. Letting $s_{kj}
=\sup\{\tilde{\l}(z-r(o)), z\in
S_{kj} \}$, we obtain
\begin{align*}
 &\eta^{-2}\int_{\l(x-r(o))<1}\rho^{(2)}(x)\l(x-r(o))\d x \\
 &\leq \eta^{-2}\sum_{(k,j)\in
 \Z^2}s_{kj}\int_{S_{kj}}\rho^{(2)}(x)\d x\\
 &=\sum_{(k,j)\in \Z^2}s_{kj} \mathcal{K}_\eta(S_{kj})\\
 &\stackrel{(a)}{<}C_{S_1}\sum_{(k,j)\in \Z^2}s_{kj}
 \stackrel{(b)}{<} \infty,
\end{align*}
where $(a)$ follows from Assumption B.1 and $(b)$ follows since $s_{kj}$ decays
like $(k^2+j^2)^{-\alpha/2}$.
Now considering the other term in \eqref{eq:sam},
\begin{align*}
  \eta^{-1}\EP[\I_1] &= \eta^{-1} \EP \Phi_t(B(r(o),b))\\
&\stackrel{(a)}{=}\mathcal{K}_\eta(B(r(o),b))\\
&\stackrel{(b)}{<} C_{B(r(o),b)} <\infty,
\end{align*}
where $(a)$ follows from Definition \ref{def:secorder}, and $(b)$ follows from \eqref{eq:ppd} and Assumption B.1.
Hence it follows from \eqref{eq:sam} that
\[\lim_{\eta \rightarrow 0}\frac{1-\p}{\eta} <
\frac{\EP[\min\{1,C\I\}]}{\eta} <\infty,\]
proving the theorem.

\section{Proof of Theorem \ref{thm:upper1} }
\label{app:proof13}
It suffices to prove that $1-\p = \Omega(\eta^{\nu\alpha/2})$. The success
  probability is
  \begin{align*}
  \p &= \PP(\ps > \frac{\T}{\l(R)} \sum_{\z \in
\Phi_t}\h{\z r(o)}\l(\z-r(o)))\\
  &\stackrel{(a)}{\leq} \PP(\ps > \frac{\T}{\l(R)} \sum_{\z \in \Phi_t\cap
  B(o,R_1\eta^{-1/2})}\h{\z r(o)}\l(\z-r(o)))\\
  &\stackrel{(b)}{\leq}\PP(\ps > \frac{\T}{\l(R)} \l(2R_1\eta^{-1/2}) X),
  \end{align*}
  where $X= \sum_{\z \in \Phi_t\cap
  B(o,R_1\eta^{-1/2})}\h{\z r(o)}$. $(a)$ follows since we have truncated the interference, and  $(b)$ follows from the fact that
  $B(r(o),R_1\eta^{-1/2})\subset B(o,2R_1\eta^{-1/2})$ for small $\eta$.
  Since $\h{\z r(o)}$ are independent random variables, $X$ can be alternatively
  written as $X = \sum_{i=1}^n \h{i}$,
  where $\h{i}$ are  i.i.d and distributed as $\h{zr(o)}$, where $n$ is the
  number of points of $\Phi_t$ in the ball $B(o,R_1\eta^{-1/2})$.
  Hence
  \[\p \leq \EP\F\left(\frac{\T}{\l(R)} \l(2R_1\eta^{-1/2}) X\right).\]
  Since $\F(x) <1$, by the dominated convergence theorem \cite{folland} and the
  expansion of $\F(x)$ for small $x$,
  \[\lim_{\eta \rightarrow 0} \frac{1-\p}{\eta^{\nu \alpha/2}} \geq
  \lim_{\eta\rightarrow 0} \eta^{-\nu \alpha/2} \EP\Theta\left(
  \left(\l(2R_1\eta^{-1/2}) X   \right)^{\nu}  \right).\]
  Using the fact that $y^\nu, y>0$ is a convex function for $\nu\geq1$, we have
 \[\lim_{\eta \rightarrow 0} \frac{1-\p}{\eta^{\nu \alpha/2}} \geq
  \lim_{\eta\rightarrow 0} \eta^{-\nu \alpha/2} \Theta\left(
  \left(\l(2R_1\eta^{-1/2}) \EP[X]   \right)^{\nu}  \right).\]
  In the limit $\l(2R_1\eta^{-1/2})^{\nu} \eta^{-\nu\alpha/2}$ tends to a
  positive constant and hence it suffices to prove that $\EP[X]^{\nu} >0$. From the
  definition of $X$ we have $\EP[X] = \EP[n]\E[\h{}]$, and hence
from  Assumption B.2, we obtain
  \[ \lim_{\eta\rightarrow 0}\EP[n] = \lim_{\eta\rightarrow 0}\eta
  K(R_1\eta^{-1/2}) >0,\]
  hence verifying our claim.

\section{Proof of Theorem \ref{thm:aloha}}
  \label{app:aloha}
\begin{IEEEproof}
Let $\I_R \triangleq \sum_{\x \in \Phi\cap
  B(o,R)} \h{\x r(o)}\l(\x-r(o))\i(\x \in \Phi_t) $. By the continuity of $\F(x)$  and the
  dominated convergence theorem, we have
  \[\EP[\F(\I)]= \lim_{R\rightarrow \infty}\EP[\F(\I_R)].\]
  Hence we have
  \[\lim_{\eta \rightarrow 0}\frac{1-\p}{\eta}= \lim_{\eta \rightarrow
  0}\lim_{R\rightarrow \infty}\frac{1-\EP \F(\I_R)}{\eta}.\]
We now show that the limits in the above equation can be exchanged, \ie,
\[\lim_{\eta \rightarrow
  0}\lim_{R\rightarrow \infty}\frac{1-\EP \F(\I_R)}{\eta} = \lim_{R\rightarrow \infty}\lim_{\eta \rightarrow
  0} \frac{1-\EP \F(\I_R)}{\eta}.\]
  For the limits to be exchanged, we require $1-\EP \F(I_R)$ should be uniformly
  convergent as $R\rightarrow \infty$.
  Define $\bar{I}_{R}$  by the relation $\I = \I_R + \bar{\I}_{R}$.
  By the Lipschitz property of $\F(x)$ we obtain
  \[\left| \EP\left[\frac{\F(\I_R)-\F(I)}{\eta}\right]\right|< C
  \frac{\EP[\bar{I}_R]}{\eta}.\]
  By the definition of the second order product density, we have
  \[\EP[ \bar{I}_R] = \eta \int_{B(o,R)^c}\rho^{(2)}(x)\l(x-r(o))\d x,\]
  and hence
 \[\left| \EP\left[\frac{\F(\I_R)-\F(I)}{\eta}\right]\right|< C
 \int_{B(o,R)^c}\rho^{(2)}(x)\l(x-r(o))\d x,\]
 and we observe that the RHS tends to zero at a rate that does not depend on
 $\eta$. Hence $(1-\EP \F(\I_R))/\eta$  tends uniformly to
 $(1-\EP\F(\I))/\eta$, and the limits can be interchanged.
   Since we are considering only simple point processes the cardinality
  $|\Phi\cap B(o,R)| <\infty$ almost surely. Define $M :\equiv |\Phi\cap
  B(o,R)|$. Hence for a finite $R$,
 \begin{align*}\EP \F(\I_R) &= \EP\sum_{k=0}^M\frac{ \eta^{k}(1-\eta)^{M-k}}{k!}\\
 &\cdot\sum_{1\leq i_1,\dots,i_k\leq M}^{\neq} \F(Y_1+\dots+Y_k), \end{align*}
 and thus
 \[\lim_{\eta \rightarrow o} \frac{1-\EP
 \F(\I_R)}{\eta}=\EP\sum_{i=1}^M(1-\F(Y_i)).
 \]
 So by the interchangeability of limits,
 \[\lim_{\eta \rightarrow 0}\frac{1-\p}{\eta} = \EP\sum_{\x \in \Phi}[1-
 \F(\h{}\l(\x-r(o))], \]
 which by the definition of the second-order product density is
 \[\lim_{\eta \rightarrow 0}\frac{1-\p}{\eta} = \int_{\R^2}
 \left[1-\E\F(\h{}\l(x-r(o)))\right]\rho^{(2)}(x)\d x. \]
\end{IEEEproof}

\section{Proof of Lemma \ref{lem:prod}}
\label{app:prod}
\begin{IEEEproof}
Let $\BB(x_1,\dots,x_n)$ denote the volume of the intersection
of discs of radius $\a$ centered around $x_i$ with the convention $\BB(x_i)=\pi
\a^2$. Let $x_k$ denote the origin $o$.  Let
\[f(m_1,\dots,m_k)= \exp\left(-\sum_{J\subset\{1,\dots,k \}} (-1)^{|J|+1}m_{\min
 J} \BB( J)\right),\]
where $\BB(J)= \BB(x_{a_1}\dots x_{a_{|J|}})$, $J=\{a_1,\dots,a_{|J|}\}$.
The $k$-th order product density is \cite{stoyan,BaccelliNOW}
\begin{align*}\rho^{(k)}_\a(x_1,\dots,x_{k-1})&= k!\int
  f(m_1,\dots,m_k)\i(0\leq m_1\leq\dots\leq m_k\leq 1)\d m_1..\d m_k.
\end{align*}
We also have
\[\BB(\a x_1, \dots, \a x_m) = \a^{2}\BBB(x_1,\dots, x_m),\]
and hence
 \begin{align*}
  \rho^{(k)}_\a(\a x_1,\dots,\a x_{k-1})
  & = k!\int  \exp\left(-\a^2\sum_{J\subset\{1,\dots,k \}} (-1)^{|J|+1}m_{\min
 J} \BBB( J)\right)\\
  &\cdot\i(0\leq m_1\leq\dots\leq m_k\leq 1)\d m_1\dots\d m_k.
\end{align*}
Using the substitution $\a^{2}m_i \rightarrow m_i$ we obtain
 \begin{align*}
   \rho^{(k)}_\a(\a x_1,\dots,\a x_{k-1})
&   =\a^{-2k} k!\int
  \exp\left(-\sum_{J\subset\{1,\dots,k \}} (-1)^{|J|+1}m_{\min
 J} \BBB( J)\right)\\
  &\cdot\i(0\leq m_1\leq\dots\leq m_k\leq \a^2)\d m_1\dots\d m_k.
\end{align*}
Using the dominated convergence theorem, the lemma is proved with
 \begin{align}
    \tilde{\rho}^{(k)}( x_1,\dots, x_{k-1})
& = k!\int
  \exp\left(-\sum_{J\subset\{1,\dots,k \}} (-1)^{|J|+1}m_{\min
 J} \BBB( J)\right)\nonumber\\
  &\cdot\i(0\leq m_1\leq\dots\leq m_k<\infty)\d m_1\dots\d m_k.
\end{align}
 \end{IEEEproof}

 \section{Proof of Theorem \ref{thm:matern}}
   \label{app:csma}
\begin{IEEEproof}
  From \eqref{eq:ccdf}, we observe that
  \begin{eqnarray*}
	\p &=& \EP \F\left(\frac{\T}{\l(R)}\I\right)\\
	&=& 1-c_0\EP
	 \left(\frac{\T}{\l(R)}\I\right)^{\nu}+\EP\left[\sum_{k=1}^{\infty}\frac{c_k}{k!}\left(\frac{\T}{\l(R)}\I\right)^{k+\nu}\right],
  \end{eqnarray*}
 So it suffices to prove that
\[\EP\left[\sum_{k=1}^{\infty}\frac{c_k}{k!}\left(\frac{\T}{\l(R)}\I\right)^{k+\nu}\right]=o\left(\EP
	\left(\I\right)^{\nu}\right)\]
	as $\eta \rightarrow 0$. Without loss of generality and for notational
	convenience we assume $\T/\l(R)=1$.   
  We have
  \[\lim_{\a \rightarrow
  \infty}\frac{\p-(1-\EP[\I^{\nu}])}{\a^{\alpha\nu}}=\lim_{\a\rightarrow
  \infty}
  \sum_{k=1}^{\infty}\frac{c_k}{k!}\frac{\EP[\I^{k+\nu}]}{\a^{\alpha\nu}}\]
  We now prove that there exists some $D_n$ such that
  $\frac{\EP[\I^{k+\nu}]}{\a^{\alpha\nu/2}}< D_k$, where $D_k$ does not depend on $\a$ and
  \[\sum |D_k  c_k|/k! <\infty.\] Then using the dominated convergence theorem, we
  can exchange the limit operation and the summation.
   Let
  $\mathcal{P}(n)$ denote the set of partitions of $n$, \ie, the set of integer
  sets
  $(p_1,
  p_2\dots,p_k)$, such that $\sum p_i = n$ and $1<p_i \leq n$. Tuples with just
  different ordering are not distinguished.Then it can be
  shown that
  \begin{align}
	\eta \EP[\I^n]= {}&\sum_{ (p_1,\dots,p_k)\in \mathcal{P}(n)}
  \int
  \rho^{(k+1)}_a(x_1,\dots,x_k)\nonumber\\
  &\cdot\prod_{i=1}^{k}\l(x_i)^{p_i} \E[\h{}^{p_i}]\d
  x_i.
  \label{eq:rho}
\end{align}
  Since
  $\rho^{(k+1)}(x_1,\dots,x_k) <\prod \i(\|x_i\|>\a)$ we obtain,
  \[ \eta\EP[\I^n]=\sum_{(p_1,\dots,p_k)\in \mathcal{P}(n)}
  \int\prod_{i=1}^{k}\l(x_i)^{p_i} \E[\h{}^{p_i}] \i(\|x_i\|>\a)\d
  x_i,\]
  which equals
  \begin{align*}
	\eta\EP[\I^n]= {} &\sum_{(p_1,\dots,p_k)\in \mathcal{P}(n)} (2\pi)^{k}\prod_{i=1}^{k}
  \E[\h{}^{p_i}] \frac{\a^{-p_i\alpha+2}}{p_i\alpha-2}\\
  ={}&\sum_{\{p_1,\dots,p_k\}\in \mathcal{P}(n)} (2\pi)^{k} \a^{-n\alpha+2k}\prod_{i=1}^{k}
  \frac{\E[\h{}^{p_i}] }{p_i\alpha-2}\\
  \stackrel{(a)}{\leq}{}&\sum_{(p_1,\dots,p_k)\in \mathcal{P}(n)}
  \left(\frac{2\pi}{\alpha-2}\right)^{k} \a^{-n\alpha+2k}b^{n} n!,
\end{align*}
where $(a)$ follows from the inequality $p_i\alpha -2 > p_i(\alpha -2)$,
 $p_i\geq 1$,
Assumption A.5 and the inequality $(n+m)! > n!m!$.
It is a known result that the cardinality of the partitions
$|\mathcal{P}(n)|\sim \exp(\pi \sqrt{2n/3})/(4n \sqrt{3})= g_n$ for large $n$
\cite{rademacher1938partition}.
Hence it follows that
\[	\eta\EP[\I^n] =O\Big( g_n
\max\left\{\left(\frac{2\pi}{\alpha-2}\right)^{n},1\right\}
\a^{-n\alpha+2n}b^{n} n!\Big),\]
as $n\rightarrow \infty$.
Let $1/\epsilon \in (0,\alpha-2)$ and for notational convenience we assume
the worst case, $\frac{2\pi}{\alpha-2}>1$.
Choose $\a=  e \left(\frac{2\pi b}{\alpha-2}\right)^{\epsilon} $. Then
\[	\eta\EP[\I^n] =O\Big(\underbrace{ e^{\pi\sqrt{2n/3}-n\alpha+2}\left(\frac{2\pi
b}{\alpha-2}\right)^{-n(\epsilon(\alpha-2)-1)} n!}_{D_n}\Big), \]
as $n\rightarrow\infty$.
Hence from Assumption A.1  it follows that
\[\sum |D_n c_n|/n!  <\infty.\] Hence using dominated convergence theorem,
\begin{equation}\lim_{\a\rightarrow
  \infty}
  \sum_{k=1}^{\infty}\frac{c_k}{k!}\frac{\EP[\I^{k+\nu}]}{\a^{\alpha\nu}}=\sum_{k=1}^{\infty}\frac{c_k}{k!}\lim_{\a\rightarrow\infty}\frac{\EP[\I^{k+\nu}]}{\a^{\alpha\nu}}.
  \label{eq:thm}
\end{equation}
  From Lemma \ref{lem:prod}, it can be observed that
  \[ \rho^{(k+1)}_\a(\a x_1,\dots,\a x_k) \sim \a^{-2(k+1)}
  \tilde{\rho}^{(k+1)}(x_1,\dots,x_k). \]
Using the substitution $x_i \rightarrow \a x_i$ in \eqref{eq:rho}, we obtain
\begin{align}
  \eta  \EP[\I^n]\sim {}& \sum_{(p_1,\dots,p_k)\in \mathcal{P}(n)}
  \a^{-2(k+1)}\int
  \tilde{ \rho}^{(k+1)}(x_1,\dots,x_k)\nonumber\\
  &\cdot\prod_{i=1}^{k} \a^{-\alpha p_i+2}\|x_i\|^{-\alpha p_i} \E[\h{}^{p_i}]\d
  x_i.
\end{align}
Hence we obtain,
\begin{align*}
  \lim_{\a \rightarrow 0}\frac{\EP[\I^n]}{\a^{\alpha\nu}} \sim &{} \lim_{\a \rightarrow 0}\pi\a^{\alpha (n-\nu)}\sum_{(p_1,\dots,p_k)\in \mathcal{P}(n)}
  \nonumber\\
&  \int  \tilde{ \rho}^{(k+1)}(x_1,\dots,x_k)
\prod_{i=1}^{k} \|x_i\|^{-\alpha p_i} \E[\h{}^{p_i}]\d
  x_i,
\end{align*}
which is equal to zero for $n>\nu$. Hence
\[\lim_{\a\rightarrow\infty}\frac{\EP[\I^{k+\nu}]}{\a^{\alpha\nu}}=0, \quad k >
0.\]
Using \eqref{eq:thm} the theorem is proved.

\end{IEEEproof}

\section{Proof of Lemma \ref{lem:approx}}
\label{app:asym}
\begin{IEEEproof}
  We first observe that as $\epsilon \rightarrow 0$, $f^*(\epsilon)\rightarrow
  0$.
  We begin by  proving the case when $f(x) > f_1(x)$ in the neighborhood of
  zero. See Figure \ref{fig:proof12} for the rest of the
  proof.
  \begin{figure}[ht]
	\begin{center}
	  \includegraphics[width=3.2in]{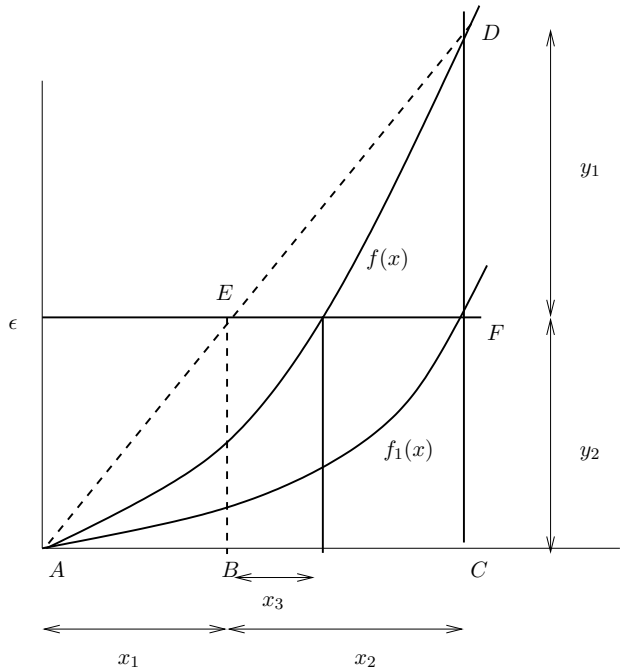}
	\end{center}
	\caption{By the local convexity of the function $f_1(x)$, 
	it follows that $f_1(x)$ lies below the line $AD$. For small $x$,
	$y_2=f_1(x)$ and $y_1 =o(f_1(x))$, \ie, $y_1/y_2\rightarrow 0$. We require to
	prove that $(x_2-x_3)/(x_1+x_3) \rightarrow 0$, which is true if
	$x_2/x_1\rightarrow 0$. But  by the congruency of triangles $ABE$ and $ACD$
	and the congruency of the triangles $DEF$ and $DAC$ it is easy to show that
	$y_2/y_1=x_1/x_2$, hence proving the claim.}
	\label{fig:proof12}
  \end{figure}
The case when $f_1(x)>f(x)$ around $x=0$  can be dealt with in a similar way.
\end{IEEEproof}

\section{Proof of Theorem \ref{thm:main}}
\label{app:main}
In this section we begin with a few lemmas which are required to prove Theorem \ref{thm:main}.

\begin{lemma}
  \label{lemm:two}
  For any stationary process $\Phi_t$,
  if
    \[
  \frac{\eta^{-1}\int_{\R^2}\rho^{(2)}_\eta(x)\Delta^{2}(x)\d x}{\k} >2, \]
 then
  \[\frac{\mathcal{G}[\exp(-\Delta(x))]}{1-\m+\k/2
  } > 1.\]
\end{lemma}
\begin{IEEEproof}
  We have
\begin{align*}
\mathcal{G}[\exp(-\Delta(x))]&= \EP\left[\prod_{\x \in \Phi_t}1- (1-
  \exp(-\Delta(x)))\right]\\
  &\stackrel{(a)}{>} 1-\EP\sum_{\x \in \Phi_t} 1-\exp(-\Delta(\x))\\
  &=1-\eta^{-1} \int_{\R^2} \rho^{(2)}_\eta(x) (1-\exp(-\Delta(x)))\d x,
\end{align*}
where $(a)$ follows from $\prod 1- a_i  \geq 1-\sum a_i$. Using the expansion
of $\exp(-x)$, we have $\mathcal{G}[\exp(-\Delta(x))]$ is greater than
\[1-\m + \sum_{m=2}^{\infty}\frac{ (-1)^{m}\eta^{-1}}{m!}\int_{\R^2}
\rho^{(2)}_\eta(x)
\Delta^m (x)\d x . \]
So it is sufficient to prove that  the summation is greater than $\kappa$.
By the inequality
\[(x-1)+\exp(-x) \geq \frac{x^2}{4}, \quad x\in [0,1],\]
it is sufficient to show
\[  (4\eta)^{-1}\int_{\R^2}\rho^{(2)}_\eta(z) \Delta^{2}(z)\d z \geq \k/2.\]
which follows from the assumption.
\end{IEEEproof}
We now show that for any  positive density of transmitters $\eta >0$, the
success probability is strictly less than one.
\begin{lemma}
  \label{lem:max}
  \[\bar{\eta} = \sup\left\{\eta \geq 0: 1-\mathcal{G}[\exp(-\Delta(x))] < \epsilon \right\}\]
 tends to zero as $\epsilon \rightarrow 0$.
  \label{lem:three}
\end{lemma}
\begin{IEEEproof}
 From the definition of the reduced probability generating functional,
 \begin{align*}
 \mathcal{G}[\exp(-\Delta(x))] &= \EP \exp(-\sum_{\x \in \Phi_\eta} \Delta(x))\\
 &\stackrel{(a)}{\geq}\exp(-\m)
 \end{align*}
 where $(a)$ follows from Jensen's inequality.
 So it is sufficient to prove that $\sup \eta$ with the constraint  $\exp(-\m) >
 1-\epsilon$ tends to zero as   $\epsilon \rightarrow 0$. But since $\m$ is the
 average of $\sum_{\x \in \Phi_t} \Delta(x)$ with respect to the Palm
 distribution and since $\Delta(x)>0$,  a necessary condition for
 $\m\rightarrow 0$ is for the density $\eta$ to tend to zero.
\end{IEEEproof}
We now return to the proof of Theorem \ref{thm:main}.
From \eqref{eq:TCineq}, it suffices to prove
	$\TCU(\epsilon)=\TCL(\epsilon)+o(\TCL(\epsilon))$. The upper bound
	$\TCU(\epsilon)$  is equal to the supremum value of $\eta$
	so that
	\[\min\{1-\m +\k/2, \mathcal{G}(\exp(-\Delta(x)))\} >1-\epsilon.\]
	This condition implies $\eta\rightarrow 0$ as $\epsilon \rightarrow 0$ by
	Lemma \ref{lem:max}.
Since $\eta$ should
be small, it follows from Lemma \ref{lemm:two} that
\[\min\{1-\m +\k/2, \mathcal{G}(\exp(-\Delta(x)))\} = 1-\m +\k/2.\]
\begin{figure}
  \begin{center}
	\includegraphics[width=3in]{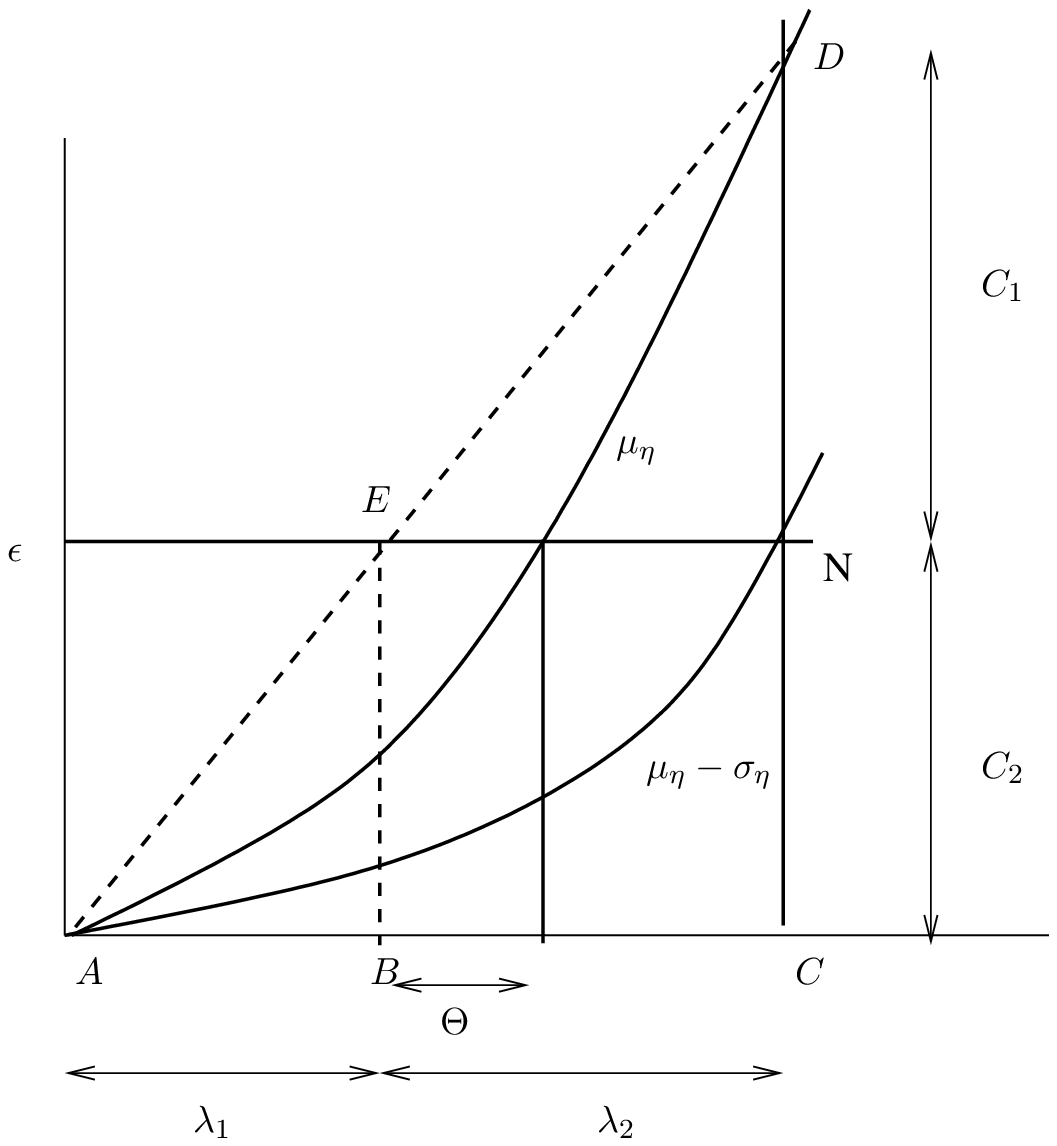}
  \end{center}
  \caption{ Proof for Theorem \ref{thm:main}. Observe that the triangle $ABE$ is
  congruent to the triangle $ACD$.}
  \label{fig:one}
\end{figure}
So the upper bound translates to finding the maximum $\eta$ such that
$\m -\k/2 < \epsilon$. Also  by our Assumption A.2,  $\m$ is locally convex in the neighborhood of
$\eta=0$. See Figure \ref{fig:one} where the upper and lower bounds are
illustrated. It suffices to prove
$\lim_{\eta\rightarrow 0}(\lambda_2-\Theta)/(\lambda_1+\Theta) =0$. Also Assumption
A.3, implies $\lim_{\eta\rightarrow 0} C_1/(C_1+C_2)=0$. Hence by the
congruency of the triangles $ABE$ and $ACD$, it follows that
$\lambda_2/\lambda_1$ tends to zero, which proves the theorem.
\bibliographystyle{ieeetr}
\bibliography{low_outage}

\end{document}